\documentclass{article}

\usepackage[english]{babel}

\usepackage[letterpaper,top=2cm,bottom=2cm,left=3cm,right=3cm,marginparwidth=1.75cm]{geometry}
\usepackage{makecell}
\usepackage{multirow}
\usepackage{booktabs}
\usepackage[table]{xcolor}
\usepackage{colortbl}
\usepackage{amsmath}
\usepackage{graphicx}
\usepackage{subcaption} 
\usepackage[colorlinks=true, allcolors=black]{hyperref}
\usepackage[backend=bibtex, style=ieee]{biblatex}
\usepackage[nocomment, commandnameprefix=ifneeded, final]{changes}
\usepackage{float}
\usepackage{adjustbox}

\title{DCGen \deleted{1.0} \added{1.1} Technical Report: Generating Datacenter Configurations (including IT, Power, Cooling)  }

\author{
\begin{minipage}{0.48\textwidth}
\centering
Wedan Emmanuel Gnibga \\
University of Chicago \\
Chicago, Illinois, USA \\
\texttt{wgnibga@uchicago.edu}
\end{minipage}
\hfill
\begin{minipage}{0.48\textwidth}
\centering
Andrew A. Chien \\
University of Chicago and Argonne National Laboratory \\
Chicago, Illinois, USA \\
\texttt{aachien@uchicago.edu}
\end{minipage}
}

\date{\vspace{1cm} \large Version 1.0
\\ \vspace{0.3cm}\today}

\begin{document}
\maketitle

\begin{abstract}




\added{Diversification of digital applications and workloads has driven the development of diverse datacenter architectures on ever-larger scales.  These datacenters consist of complex IT, power, and cooling systems with interdependencies that influence configuration and performance. As datacenters scale and power density increase, designing realistic models becomes more difficult, particularly for research, because it requires understanding all layers of the datacenter and how they interact. Consequently, many studies rely on outdated or unrealistic designs.} \deleted{The increasing demand for digital services has led to a proliferation of diverse datacenter architectures.} 

To support research in datacenter hardware design principles, operational dynamics, cooling mechanisms, and interactions of these facilities with the electrical grid, we have designed \added{\textbf{DCGen}}, a tool which
can generate a variety of datacenter configurations (including IT hardware, cooling and power distribution infrastructures)
at various electrical power, compute capability, and area targets. \added{The tool captures power and space characteristics of IT, cooling, and power infrastructures at both the rack and datacenter levels, enabling modeling of power, energy, and space.}

\deleted{We first performed a survey of reference computing
servers and rack-level systems designed for a variety of purposes.  These reference systems are analyzed, and used to create models for canonical capability, power, and density of racks in a datacenter (eg. AI training rack).  These racks can then be used by the model to create a variety of  datacenter types. 
}

\deleted{W\added{Then, w}e developed \textbf{DCGen}, a model-driven tool which can generate a variety of representative datacenter hardware configurations at various electrical power, compute capability, and area targets.} DCGen leverages specific use cases such as AI training, AI inference, and cloud services, to select reference and canonical IT hardware configurations, producing realistic mixes of server types. It can target datacenter scale in terms of \deleted{either} \added{both} power (e.g., 10 MW, 100 MW, 1 GW) \deleted{or}  \added{and} compute capability. For cooling and  power distribution infrastructures, DCGen chooses components from a production equipment catalog that optimizes for space or power efficiency while meeting the datacenter capacity requirements.
\deleted{This tool enables a wealth of research using realistic datacenter designs through "what-if" scenario exploration}
\added{This tool supports research using realistic datacenter designs through ``what-if'' scenario exploration, including studies of power density evolution over time, grid interconnection capacity planning, datacenter-grid interactions, and space management. }

\added{DCGen is available as open-source software at the following link}\footnote{\added{DCGen source code:} \href{https://github.com/WedanEmmanuel/DCGen}{\added{https://github.com/WedanEmmanuel/DCGen}}\label{DCGen-git-link}.}.
\end{abstract}

\vspace*{0.5in}

\noindent {\large \textbf{Subcontractor:} University of Chicago} \\
{\large \textbf{NLR Subcontract number:} SUB-2025-10203} \\
{\large \textbf{Subcontract Monitor:} Guangdong Zhu, guangdong.zhu@nrel.gov} \\
{\large \textbf{Deliverable Date:} May 14, 2025} \\
{\large \textbf{Deliverable Description:} 6.1, Datacenter Canonical Configurations} \\

\newpage
\tableofcontents 

\newpage
\section{Introduction}
\deleted{In today’s digital world, datacenters serve as the backbone of almost every online service, from streaming and video conferencing platforms to Cloud and artificial intelligence applications. At the heart of these facilities lies a complex arrangement of physical hardware components that vary depending on the type of workload they serve and their electrical power.
That refers to the deployment and integration of servers, storage systems, networking devices, power systems, and cooling infrastructure, all of which work together to ensure high availability and performance. The effectiveness of a datacenter depends largely on how well its hardware is organized and optimized to support current workloads and future growth.}

\deleted{Datacenter hardware falls into three categories: (1) IT hardware(servers, racks, network devices, data storage, etc.), (2) cooling system, and (3) power system. This document focuses on the description of IT hardware configurations 
First, we catalog datacenter use cases (from Cloud to AI oriented datacenters), using them to create compute and power models for canonical racks.  Then we detail some base datacenter IT\added{, cooling and power distribution} configurations.}

\deleted{We build \textbf{DCGen}, a model-driven tool, out of these reference datacenters. This tool automates \deleted{the IT} \added{datacenter} configuration process for a target datacenter specified by its number of racks or its electrical power.  Moreover, DCGen facilitates the analysis datacenters power density and spatial footprint.  Finally, it enables \added{to} make projections and analysis of future datacenters.}


\added{Datacenter workloads are increasingly diverse, including cloud services, high-performance computing, AI training, inference, etc. Each of these workloads has different requirements on compute, memory, storage, and networking, introducing significant diversity in IT hardware requirement, the resulting thermal behavior, and power consumption across time. The physical infrastructure that supports these workloads is composed of tightly interconnected subsystems spanning IT racks, power distribution, and cooling.  These interdependencies are both static and dynamic. Statically, they constrain design choices. Server architectures and rack power densities shape the design of power distribution infrastructures and influence cooling requirements, while cooling system choices (e.g., air or liquid cooling, evaporative or dry cooling) determine the available cooling capacity and power demand. Dynamically, workload placement, utilization levels, and weather conditions vary over time, introducing additional complexity in cross-layer interactions. These complex static and dynamic interdependencies determine appropriate datacenter configurations, and affect performance and operational efficiency.}

\added{The rapid increase of workload heterogeneity, datacenter scales, power density, and performance requirements has therefore made realistic datacenter design a complex task  for research communities. Modern facilities may operate at hundreds of megawatts to gigawatts, with compute racks that can exceed tens or even hundreds of kilowatts. Designing and analyzing such systems requires reasoning across multiple layers (workload characteristics, IT rack layouts, power distribution, and cooling infrastructure). Therefore, it requires detailed knowledge of both IT systems and facility infrastructure, as well as their cross-layer interactions. As a result, many research studies rely on simplified, outdated~\cite{barroso2019datacenter}, or unrealistic datacenter designs.}

\added{
To solve this problem, we created \textbf{DCGen}, a model-driven tool that generates realistic datacenter designs that each include full IT, power, and cooling systems.  Because datacenters are increasingly specialized to workload, DCGen generates configurations for  four datacenter types: AI Training, mixed AI Training and Inference, AI Inference, and Cloud.  For each type, datacenters of different scales can be generated (e.g., 1 MW, 100MW, 10GW, etc.).  DCGen is a datacenter design  and model generator that makes the following key contributions:}

\begin{itemize}
    \item \added{It captures rack- and datacenter-level properties of computer servers, networking, etc, enabling a representative modeling of power and energy dynamics. The properties are: rack arrangements per node type (e.g., GPUs, storage…), rack power density, IT power capacity, and “white space” (or IT space) requirements}
    \item \added{It models cooling and power distribution hardware setup optimized for space or power, and the associated power demand and “gray space” (or non-IT space) requirement, enabling systematic power-efficiency and space management analysis.}
\item \added{It provides projections of future datacenter configurations (2027, 2029), allowing modeling and study of future datacenter systems}
\end{itemize}

\added{
The tool can generate configurations based on two parameters: (1) a specific number of racks (representing the desired compute capability), and (2) a target electrical power use. DCGen  takes as input a target compute capability or power capacity, one of four canonical datacenter types (AI Training, Mixed Training and Inference, AI Inference, and Cloud), the date of operation (2024, 2027, or 2029), the specification of optimization objective (space or power), redundancy and safety margins for cooling and power systems. It then generates the datacenter configuration (including IT, cooling and power systems). Moreover, DCGen provides detailed metrics such as rack power density, electrical loads, and space requirements\footnote{This space requirement represents an estimate of the minimum area needed to deploy the IT, cooling, and power infrastructure. It does not represent the total building footprint, which may be constructed to accommodate future expansion or repurposed from different use cases (e.g., xAI Colossus 1 building~\cite{xAIColossus}). Consequently, the actual building may include infrastructure supporting other use cases, unused space, and areas reserved for future expansion. The total building footprint of the datacenters we examined can be up to 500\% larger than our minimal space-optimized design estimate. However, limited public information on the allocation and utilization of this additional “gray space” makes it difficult to identify the specific factors contributing to this difference.\label{footnote_space}} as described above, enabling a wealth of research.}
\added{
As datacenters scale toward gigawatt-level power consumption, DCGen enables a broad range of research using realistic datacenter designs. This includes studies of power demand forecasting, cooling and power distribution system design under increasing rack densities,  the systemic impact of current and future large-scale datacenters on local and regional power grids, sustainability analysis, and renewable energy integration strategies. It is worth noting that DCGen focuses on hardware configuration generation rather than operational simulation. The datacenter designs output generated by DCGen are JSON-encoded, incorporate canonical IT configurations and commercial “gray space” components, and can be directly input into other simulation tools for studies of datacenter operation. 
}

\added{
In this document, we first catalog datacenter use cases (from Cloud to AI oriented datacenters), using them to create compute and power models for canonical racks (Section~\ref{Sec:DatacenterUseCases}). We then conduct an extensive survey of datacenter systems and nodes that support AI datacenter and cloud workloads, as described in Section~\ref{Sec:REFERENCE-DESIGNS}. Next, Section~\ref{Sec:Modeling} presents the models implemented by DCGen for generating IT, cooling, and power systems for target datacenters. Finally, Section~\ref{Sec:CaseStudies} provides exemplar case studies based on the tool.
}

\section{Datacenter Use Cases}
\label{Sec:DatacenterUseCases}
\begin{figure}[H]
    \centering
    \begin{subfigure}[b]{0.47\textwidth}
        \includegraphics[width=0.95\textwidth]{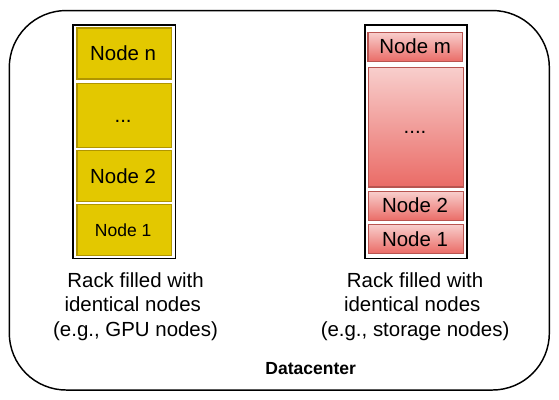}
        \caption{Arrangement of nodes and racks in a datacenter.}
        \label{fig::DatacenterLayout}
    \end{subfigure}
    \begin{subfigure}[b]{0.47\textwidth}
        \includegraphics[width=\textwidth]{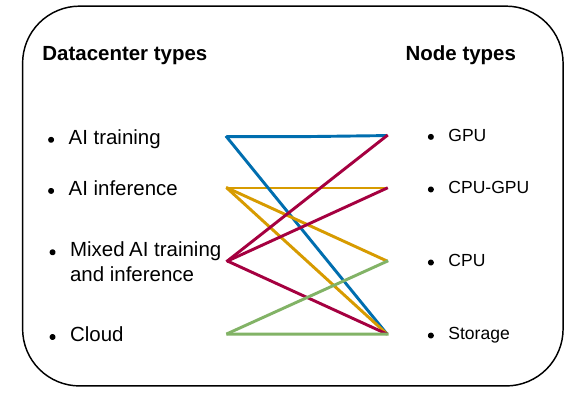}
        \caption{Links between datacenter and node types.}
        \label{fig:DataCentersLink}
    \end{subfigure}
    
    \caption{datacenter hardware configurations. The hardware setup involves identical racks, each housing only one type of node.}
    \label{fig:enter-label}
\end{figure}

In this section, we categorize datacenters based on their workload and \deleted{the associated types of nodes they utilize} \added{type of node utilized}. The arrangement of nodes and racks in a typical datacenter is depicted in Figure~\ref{fig::DatacenterLayout}. \deleted{As shown, we use the model that racks are homogeneous, meaning each rack contains only one type of node} \added{In this figure, the racks are assumed to have homogeneous nodes}.   We also generally assume that racks are full.  Both of these assumptions can be inaccurate in detail, but this assumption reflects the ideal -- as full and dense a datacenter IT configuration as possible.  
In our study, a datacenter with different node types will feature distinct groups of these homogeneous racks, all of the same physical size.
Figure~\ref{fig:DataCentersLink} relates the different datacenter types and the nodes they require.

\subsection{Server / Node Classification}

Nodes used in datacenters fall into four groups:
\begin{itemize}
    \item \textbf{GPU nodes}: \deleted{these servers} \added{they} include one or more Graphics Processing Units (GPUs). They are designed to handle highly parallel computing tasks. \deleted{It is worth noting that although these servers may contain one or a few standard CPUs, the compute capability and electricity draw of the GPUs are dominant.}  \added{They may contain one or a few standard CPUs, but the compute capability, power draw and cooling requirements of the GPUs are dominant.}
    \item \textbf{CPU-GPU nodes}: \deleted{these servers} \added{they} include one or more GPUs/accelerators in addition to the Central Processing Units. The GPUs in these category may be less compute intensive, less numerous, and/or from older generations than in GPU nodes. 
    \item \textbf{CPU nodes}: these are compute nodes that contain only CPUs (no GPUs) for general-purpose processing tasks.
    \item \textbf{Storage nodes}: their main role is to store and manage data. \deleted{Unlike compute nodes (CPU or GPU nodes), storage nodes are optimized for I/O performance and capacity rather than raw processing power. In fact, they feature a few CPUs for handling data processing tasks, but their primary component is a collection of storage bays capable of holding several terabytes of data.}
    \added{ Storage nodes are optimized for I/O performance and capacity. Typically they feature a few CPUs paired with a large collection of storage bays containing rotating hard disk drives (HDDs) and solid-state-disks (SSDs), capable of holding many terabytes of data. 	 }
\end{itemize}

\subsection{Datacenter Classification}
Datacenter designs vary based on the  specific range/category of workload to be supported.  We classify datacenters on this basis, as it determines both workload requirements and associated compute capability directly.  It does so by influencing the selection of appropriate IT hardware. This section outlines four distinct datacenter types.


\begin{itemize}
    \item \textbf{AI training datacenters}: they  are designed and optimized to handle the immense computational demands of training artificial intelligence (AI), machine learning (ML) models, and large-scale LLM scale models. 
    \deleted{These datacenters contain several racks of \textbf{GPU nodes} delivering substantial computation, and \textbf{storage nodes} holding the massive (often at the exabyte scale) datasets. } \added{These datacenters are dominated by dense 	racks of \textit{GPU nodes} delivering efficient AI training computation, and \textit{storage nodes} holding the massive (often at the 	exabyte scale) datasets.}
    
\item \textbf{AI inference datacenters}:  \deleted{AI inference designates the phase when a trained  model is used to predict/draw conclusions in application. 
This includes both large LLM models but also smaller models.   Some training of smaller ML models will happen in these datacenters.  This may necessitate less computation than the training phase. These datacenters contain \textbf{CPU-GPU, CPU, and storage nodes}.}
\added{ they are optimized for application of AI models (AI inference). This includes both large LLM models but also smaller models. Some training of smaller ML models will happen in these datacenters. This may necessitate less computation than the training phase. These datacenters contain \textit{CPU-GPU, CPU}, and \textit{storage nodes}.  
}

\item \textbf{Mixed AI training and inference datacenters}: \deleted{these facilities are designed to handle both the process of training AI models and running (inferencing) these models. This includes both large LLM models but also smaller models.   Some training of smaller ML models will happen in these datacenters.  Mixed AI training and inference datacenters are composed of \textbf{GPU and CPU-GPU, and storage nodes}.}
\added{these facilities are optimized for a mix to handle both training AI models and applying (inferencing) these models. This includes both large LLM models as well as smaller models (eg. SLM’s and non-LLM models). Some training of smaller ML models will happen in these datacenters. Mixed AI training and inference datacenters are composed of \textit{GPU, CPU-GPU}, and \textit{storage nodes}.}

\item \textbf{Cloud datacenters}:  traditional datacenters primarily focus on data storage and general-purpose computing. Cloud datacenters are based on racks of \textbf{CPU and storage nodes}.
Some inference will be part of the cloud workloads.
\end{itemize}

It is worth noting that in some datacenter configurations, storage devices may be local to the compute nodes or racks, thus not requiring separate storage nodes. 
\section{Survey of datacenter IT, Cooling and Power Distribution Configurations}

\label{Sec:REFERENCE-DESIGNS}

\subsection{Reference Datacenter IT configurations}
\label{SEC::BASE-DATACENTERS}
In this section, we describe the IT hardware configuration of a set of datacenters. \added{The systems analyzed span a large spectrum of real-world deployments, from hyperscale AI training clusters, to general-purpose cloud infrastructures.  We selected these datacenters based on two key criteria: (1) their deployment dates, prioritizing recent and announced future configurations, and (2) the availability of data describing their characteristics (node types, rack layout, power capacity, compute capability, etc).}
These hardware configurations may be utilized in two ways in the tool: 
\begin{enumerate}
    \item \deleted{they may be utilized as reference configuration for generating datacenters that mimic a particular IT layout.} \added{As reference configurations for generating datacenters IT configurations of varied scale}. An example of case study using one of these configurations is  shown in Section~\ref{sec:SingleDCConfig}.
    \item \deleted{they may also be aggregated into canonical hardware configurations, allowing to build datacenters that capture characteristics from multiple hardware setups} \added{As inputs for aggregation into canonical hardware configurations, allowing the generation of datacenter designs that capture characteristics from multiple hardware setups (see Section~\ref{sec:CANONICAL-IT-MODEL})}.  A detailed study is shown in Section~\ref{Sec:DC-Example}.
\end{enumerate}

\added{In the following, we present the collected characteristics for each AI-oriented (Section~\ref{Sec:AI-SURVEY}) and cloud-oriented (Section~\ref{Sec:Cloud-SURVEY}) configurations.}

\subsubsection{IT Hardware Configurations in AI Training and Inference Datacenters}

\label{Sec:AI-SURVEY}
In this section, we describe reference datacenter configurations used for AI training and/or AI inference.  

Table~\ref{tab:AIDCConfig} summarizes the key parameters of these configurations, where GPU racks,  CPU-GPU racks,  CPU racks, and  storage racks stand for the racks hosting respectively GPU, CPU-GPU, CPU, and storage nodes in each datacenter.


\renewcommand{\arraystretch}{2.75}
\begin{table}[]
    \centering
    \Large
    \resizebox{\textwidth}{!}{%
    \begin{tabular}{|c|c|c|c|c|c|c|c|c|c|c|}
    \hline
     \multirow{2}{*}{\textbf{Datacenter}} & \multirow{2}{*}{\makecell{\textbf{Compute} \\ \textbf{Capability}}} & \multirow{2}{*}{\makecell{\textbf{Rack size} \\ \textbf{(RU)}}}  & \multicolumn{2}{c|}{\textbf{ GPU racks}}   & \multicolumn{2}{c|}{\textbf{ CPU-GPU racks}} &  \multicolumn{2}{c|}{\textbf{CPU racks}} & \multicolumn{2}{c|}{\textbf{ Storage racks}}  \\
     \cline{4-11}
     &  & & \textbf{\#racks} & \makecell{ \textbf{Peak power}\\ \textbf{(kW/rack)}} 
     & \textbf{\#racks} & \makecell{ \textbf{Peak power}\\ \textbf{(kW/rack)}}
     & \textbf{\#racks} & \makecell{ \textbf{Peak power}\\ \textbf{(kW/rack)}}
     & \textbf{\#racks} & \makecell{ \textbf{Peak power}\\ \textbf{(kW/rack)}}\\
     \hline 
     \rowcolor{gray!30}  
    \parbox[c]{0.31\linewidth}{\centering El Capitan\\(2024)~\cite{ElCapitain}}
     & 2.79 EFLOPS &HPC 48U  &87 & 400 & - & - & - &  - & \multicolumn{2}{c|}{ Local storage} \\ \hline 
     
    \rowcolor{gray!30}  
     \parbox[c]{0.31\linewidth}{\centering xAI COLOSSUS\\ (2024)~\cite{xAIColossus100k}} & 200 EFLOPS FP16 & HPC 48U & 1563  & 100  & - & - &  - & - & 203 & 34 \\ \hline 

     \rowcolor{gray!30}  
     Aurora(2024)~\cite{AuroraArchi} & 2 EFLOPS & HPC 42U &  166 & 128.8 &- & - &  - & - & 64 & 11.3\\ \hline

    \rowcolor{gray!30}  
   \parbox[c]{0.31\linewidth}{\centering NVIDIA DGX \\ SuperPOD (2024)~\cite{DGXSuperPOD}} & 360 PFLOPS FP16 & 42U & 8 & 120  & - & - &  - & - & 2 & 21.1\\ \hline
    
     \rowcolor{gray!30}  
    \parbox[c]{0.31\linewidth}{\centering NVIDIA Rubin Ultra\\NVL576 (2027)~\cite{NVL576}} & 5 EFLOPS FP8 & 42U & 1 & 600 &  - & - &  - & - & 1 & 26.3\\
     \hline
     \hline
     \rowcolor{gray!30}  
    \parbox[c]{0.31\linewidth}{\centering PowerEdge R760xa \\ rack (2024)~\cite{PowerEdgeR760xa}} & 140.4 PFLOPS & 42U & - & - & 6 & 58.8 &  - & - & 1 & 29.7\\ \hline
     
     \rowcolor{gray!30}  
    \parbox[c]{0.31\linewidth}{\centering Supermicro inference \\ rack (2024)~\cite{SupermicroRack}} & 63.3 PFLOPS & 48U & - & - & 14 & 51.2 &  - & - & 1 & 34\\ \hline
    \rowcolor{gray!30}  
   \parbox[c]{0.31\linewidth}{\centering IBM’s Gen AI\\ (2024)~\cite{gershon2024infrastructure}} & 30 PFLOPS & 42U & - & - & 25 & 40 &  - & - & 1 & 29.7\\ \hline
      
     \makecell{Google GDC Edge\\ Rack(2021)~\cite{CDGHardware}} & - & 46U & - & - & 1 & 16.5 &  - & - & - & -\\ \hline
     
  \parbox[c]{0.31\linewidth}{\centering  ChatGPT \\ (2022)~\cite{ChatGPTInference}} & 18.1 EFLOPS &40U & - & - & 362 & 52 &  - & - & 26 & 29.7\\ \hline
     
    \parbox[c]{0.31\linewidth}{\centering Lenovo ThinkSystem \\SR670 V2 (2021)~\cite{ThinkSystemSR670V2}} & 2.15 PFLOPS &  42U & - & - & 348 & 33.6 &   - & - & 1 & 29.7\\ \hline

    \parbox[c]{0.31\linewidth}{\centering  Dell PowerEdge\\ XE7745 (2025)~\cite{DellPowerEdgeXE7745}} & 17.5 PFLOPS F16U rack  &  42U & - & - & 43 & 33.6 & - & - & 1 & 29.7\\ \hline
    \end{tabular}
    }
    \caption{Summary of hardware configurations used in AI training and inference datacenters. The grey cells show the most recent configurations (2024 and later).}
    \label{tab:AIDCConfig}
\end{table}

 In the following, we show details of each reference datacenter configuration.

\vspace{0.5cm} \noindent \textbf{NVIDIA Rubin Ultra NVL576~\cite{NVL576}} 
\begin{itemize}
    \item 1 Rack
    \item 5EFLOPS FP8 training  
    \item Up to 576 GPUs~\cite{NVL576GPUs}
    \item 144 NVLINK Switches (1500 PB/s) 
    \item 365TB memory 
    \item Storage estimation
    \begin{itemize}
        \item Taking example of the  708W (2x 270W of CPU + 8x21W SSD) Storage SuperServer SSG-121E-NE3X12R nodes (8 x 6.4 SSD)~\cite{SuperServerSSG121ENE3X12R,SSD_datasheet}
        \item 1x 600kW x 4.2\% / 95.8\% = 26.3kW storage = 38x 1U storage nodes = 1 rack 
    \end{itemize}

    \item Electrical power 
    \begin{itemize}
        \item Compute racks: \textbf{600kW/rack}
        \item Storage racks: \textbf{26.3kW/rack}
    \end{itemize}

    \item Announced for 2027
\end{itemize}

\vspace{0.5cm} \noindent \textbf{Frontier super computer~\cite{FrontierSupercomputer}}
\begin{itemize}
    \item 1.6 EFLOPS compute capability
    \item Workload: AI training and inference 
    \item 74 HPE Cray EX supercomputer cabinets ( 48U HPC), 
    \begin{itemize}
        \item 9,408 HPE Cray EX nodes. Each node contains:
        \begin{itemize}
            \item 1 HPC and AI Optimized AMD EPYC CPU 
            \item 4 Purpose Built AMD Radeon Instinct GPU (9,400 CPUs and more than 37,000 GPUs total)
            \item CPU-GPU interconnect: AMD Infinity Fabric
        \end{itemize}
        \item 9.2 petabytes of memory (half HBM, half DDR4)  
    \end{itemize}
    \item Storage
        \begin{itemize}
            \item 37 petabytes of node-local storage 
            \item 716 petabytes of center-wide storage
            \begin{itemize}
                \item 480+5400 NVMe SSDs based on  E1000 SSU-F devices (2U nodes)~\cite{FrontierStorageConfig}
                \begin{itemize}
                    \item 24 SSD discks per E1000, or total 245 x E1000 SSU-F 
                    \item 19 nodes per 48U rack~\cite{FrontierStorageDetail}, or 13 racks 
                \end{itemize}
            
            \item  47,700x HDD based on E1000 SSU-D devices (106 HDDs in 6U)~\cite{FrontierStorageConfig}
            \begin{itemize}
                \item  47,700/106 = 450 E1000 SSU-D devices 
               \item  9 nodes per 48U rack~\cite{FrontierStorageDetail}, or 50 racks 
            \end{itemize} 
            \end{itemize}
        \end{itemize}
        
    \item  Power demand
    \begin{itemize}
        \item 24.6MW total power demand\cite{FrontierTop500}
        \begin{itemize}
            \item Storage ~\cite{FrontierStoragePower}
            \begin{itemize}
                \item   1472W  per E1000 SSU-F device 
                \item 1728W  per E1000 SSU-D device 
                \item Total Storage: 1.472 * 245 + 1.728*450 = 1138.24kW  (1.14MW)
                \item 1.14MW/(113+13) = \textbf{18.1kW/rack}
            \end{itemize}
            \item Compute nodes
            \begin{itemize}
                \item Total: 24.6 - 1.14 = 23.46 MW
                \item 23.46/74 = \textbf{317kW/rack}
            \end{itemize}

        \end{itemize}
    \end{itemize}
    \item  Liquid cooled using warm water (85 ° F / 29.4 ° C), 6000 gallons~\cite{FrontierCooling}
    \item Deployment: Sep. 2021, completion: May 2022 
\end{itemize}

\vspace{0.5cm} \noindent \textbf{El Capitan - NNSA’s exascale machine~\cite{ElCapitain}}

\begin{itemize}
    \item 2.79 EFLOPS compute capability
    \item High-performance computing, generative AI, and ML training 
    \item Compute cabinets: 87 
    \item 11,136 compute nodes powered by the AMD Instinct MI300A Accelerated Processing Units (APUs). 4 x MI300A APU per node.  Each APU is: 
    \begin{itemize}
        \item 1 AMD CDNA3 GPU (228 GPU compute units, 912 Matrix cores)
        \item 24 AMD ‘Zen 4’ x86 (AMD EPYC™) CPU cores = 96/node  
    \end{itemize}
    \item Memory: 128 GB HBM3 
    \item 400 Gbps Ethernet or InfiniBand 
    \item Storage : Rabbit near-node storage (Installed directly into the supercomputer's chassis)~\cite{RabbitStorage} 
    \begin{itemize}
        \item Local storage nodes (4U nodes), 18 SSDs (16 and 2 spares)  per node
        \item 1 (AMD Epyc) storage processor. 
        \item 1 storage module per chassis (each chassis houses 8 blades / 16 compute nodes, so 696 total storage modules)  
    \end{itemize}
    \item Electrical power
    \begin{itemize}
        \item Peak power = 34.8 MW 
        \item Per rack: 34.8MW/87 = \textbf{400kW/rack}
        \item TDP per APU: 	550W (air \& liquid cooling), 760W (liquid cooling) 
    \end{itemize}
    \item Deployed in mid-2024 
\end{itemize}

\vspace{0.5cm} \noindent \textbf{xAI COLOSSUS~\cite{xAIColossus100k}}

\begin{itemize}
    \item Estimated compute capability: 200 EFLOPS FP16 (2PFLOP/GPU)
    \item Training and Inference on GPUs 
    \item Processing and data manipulation tasks on CPUs  
    \item GPU racks
    \begin{itemize}
        \item 100,000 NVIDIA Hopper Tensor Core GPUs  aranged in 4U servers
        \item 1x 4U servers = 8 NVIDIA HGX H100 GPUs 
        \item 1 Rack = 8 x 4U servers + Supermicro Coolant Distribution Unit (CDU) + associated hardware = 64GPUs/rack or total 1563 racks 
        \item Racks are arranged in groups of eight for 512 GPUs, plus networking 
    \end{itemize}
        
    \item  CPU-powered computer racks
    \begin{itemize}
        \item Dual Intel Xeon MAX 9480 CPUs in 1U nodes
        \item Each has a full set of 16 DDR5 DIMM slots for 32 total 
    \end{itemize}

    \item Storage: 1U NVMe Storage Nodes arranged into racks 
    \begin{itemize}
        \item  500PB for 100,000 GPUs~\cite{xAIColossus}
        \item Taking example of the  708W Storage SuperServer SSG-121E-NE3X12R nodes (8 x 6.4 SSD)~\cite{SuperServerSSG121ENE3X12R,SSD_datasheet}
        \begin{itemize}
            \item 500PB/8x6.4 = 9765 x 1U servers
            \item 203 x 48U racks 
        \end{itemize}
    \end{itemize}
    \item Storage required per TFLOPS:  500PB/200EFLOPS = 0.0025TB/TFLOPS
    
    \item 400GbE fiber Ethernet  
    \item 3.6Tbps of bandwidth per GPU compute server 
    \item Power usage 
        \begin{itemize}
            \item  GPU racks: \textbf{100kW/rack} (48U)~\cite{xAIRackPower}.  
            \item Storage: taking example of the 708W Storage SuperServer SSG-121E-NE3X12R nodes~\cite{SuperServerSSG121ENE3X12R,SSD_datasheet} : \textbf{34kW/rack} (48U)
            \item Overall, storage represents 4.2\% of the total power usage. This number will be used as reference to estimate the storage requirement in the configurations where storage data is lacking. 
        \end{itemize}
    \item Gone live on July 22, 2024	 
\end{itemize}

\vspace{0.5cm} \noindent \textbf{Aurora Exascale (Argonne national Lab)~\cite{AuroraArchi}}
\begin{itemize}
    \item Compute capability : 2 EFLOPS
    \item 166 Compute racks (HPC racks). In each rack:  
    \begin{itemize}
        \item 64 Compute blades (10624 total) 
        \begin{itemize}
            \item 2x 4th Intel Xeon Max Series 9470C w HBM (128 GB HBM, 1024 GB DDR5 per blade) 
            \item 6x Intel datacenter GPU Max Series 1550 (768 GB HBM per blade) 
            \item 8 NICs 
        \end{itemize}
        \end{itemize}
        
        \item 32 Switch blades (64 ports):  200 Gbps (Dragonfly network topology)
        \item 6 liquid-cooled PSUs/rectifiers per rack 
        \item GPU – GPU Interconnect: Xe Link 
        \item Storage
        \begin{itemize}
            \item 230 PB NVMe with 31TB/s 
            \item 64 racks (HPC racks) with total 1024 DAOS nodes (16 nodes per rack) 
        \end{itemize}

        \item Power demand 
        \begin{itemize}
            \item  Peak power: 38.7MW~\cite{AuroraPower} 
            \item Storage 
            \begin{itemize}
                \item Let us consider as reference the 708W Storage SuperServer SSG-121E-NE3X12R nodes (8 x 6.4 SSD)~\cite{SuperServerSSG121ENE3X12R,SSD_datasheet}
                \item Storage rack power = 708W * 16 nodes = \textbf{11.3kW/rack} 
                \item Total storage Power =11.3kW * 64 racks = 723.2kW  
            \end{itemize}
            
            \item Compute racks 
            \begin{itemize}
                \item Compute rack power = (38.7MW - 723.2kW)/166racks = \textbf{228.8kW/rack} 
            \end{itemize}

    \end{itemize}
    \item Fully installed on June 23, 2023 
\end{itemize}

\vspace{0.5cm} \noindent \textbf{NVIDIA DGX SuperPOD With DGX GB200 Systems\cite{DGXSuperPOD}}
\begin{itemize}
    \item 360 PFLOPS FP16
     \item Training and inferencing trillion-parameter generative AI models 
    \item 8x NVIDIA GB200 NVL72 racks. Each rack contains
    \begin{itemize}
        \item 18 compute nodes
        \item 2 GB200 Grace Blackwell Superchips per node 
        \begin{itemize}
            \item 2x  NVIDIA Blackwell GPUS  per Superchip
            \item 1x  NVIDIA Grace CPU per Superchip
        \end{itemize}
        
        \item Total
         \begin{itemize}
             \item 72 x NVIDIA Blackwell GPUS 
             \item 32 x NVIDIA Grace CPUs 
         \end{itemize}
    \end{itemize}
    \item 30 terabytes (TB) of fast memory
    \item 130 terabytes per second (TB/s) of bidirectional GPU bandwidth 
    \item Storage estimation
    \begin{itemize}
        \item Taking the example of the  708W Storage SuperServer SSG-121E-NE3X12R nodes (8x6.4TB SSD)~\cite{SuperServerSSG121ENE3X12R,SSD_datasheet}. 
        \item Storage power = 4.2\% x 8 racks x 120kW /95.8\%= 42.1kW 
        \item Servers: 42.1/708W = 60 x 1U servers = 2 x 42U racks 
    \end{itemize}
    \item Power demand
    \begin{itemize}
        \item GPU racks: \textbf{120kW/rack}
        \item Storage racks: \textbf{21.1kW/rack}
    \end{itemize}
    
    \item Announced on March 18, 2024  
\end{itemize}

\vspace{0.5cm} \noindent \textbf{ Supermicro cloud-scale inference rack~\cite{SupermicroRack}}
\begin{itemize}
    \item Estimated compute capability:  63.3PFLOPS per rack (1,979TFLOPS per NVIDIA GH200 Grace Hopper chip~\cite{NVIDIAGPUsPerf}) 
    \item 48U rack 
    \item 32x 1U inference nodes (Supermicro GPU ARS-111GL-NHR nodes~\cite{SupermicroARS-111GL-NHR})   
    \item CPU: NVIDIA 72-core NVIDIA Grace CPU on GH200 Grace Hopper™ Superchip 
    \item GPU: 1x NVIDIA H100 Tensor Core GPU on GH200 Grace Hopper™ Superchip (Air-cooled)  
    \item Local storage: 8x Hot-swap E1.S drives + 2x M.2 NVMe drives 
    \item External storage requirements:  
    \begin{itemize}
        \item Considering the same storage to compute capability ratio as in Microsoft inference datacenters ~\cite{microsoft2024nccadsh100v5}: 404 IOPS/TFLOPS 
        \item Taking example of Storage SuperServer SSG-121E-NE3X12R nodes (8 x 6.4TB SSD) 
        \item 1 x 48U storage rack = 48 x 8x900,000IOPS =  345,600,000IOPS

        \item 1 storage rack can serve up to:  345,600,000IOPS / (404 IOPS/TFLOPS ) =  14 x 63.3PFLOPS racks 
        \item 1 storage rack serves 14 compute racks.
    \end{itemize} 
    \item Electrical power 
    \begin{itemize}
        \item Peak per node: 1.6kW~\cite{SupermicroARS-111GL-NHRPower}
        \item Peak per rack: 32x 1.6= \textbf{51.2kW/rack }
    \end{itemize}
    \item Generation : 2024  
\end{itemize}

\vspace{0.5cm} \noindent \textbf{ PowerEdge R760xa nodes~\cite{PowerEdgeR760xa}}
\begin{itemize}
    \item Estimated compute capability: 140.4 PFLOPS FP16 per 42U rack.

    \item 2U rack server 
    \item GPUs: 2 options 
    \begin{itemize}
        \item  Up to 4 x 400 W DW PCIe x16  GPU cards -- NVIDIA H100 NVL (350-400W), 1671TFLOPS FP16 per GPU~\cite{NVIDIA-H100-NVL} 
        \item  Up to 12 x 75 W SW PCIe x8 GPU cards 
    \end{itemize}
    \item CPUs: 
    \begin{itemize}
        \item Up to two 4th Generation Intel Xeon Scalable processor with up to 56 cores per processor and optional Intel® QuickAssist Technology 
        \item Up to two 5th Generation Intel Xeon Scalable processor with up to 64 cores per processor and optional Intel® QuickAssist Technology 
    \end{itemize}
    \item Memory:  32 DDR5 DIMM slot, supports RDIMM 8 TB max 
    \item Storage: 
    \begin{itemize}
        \item Up to 6 x E3.S Gen5 NVMe, max 46.08 TB  
        \item Up to 6 x 2.5-inch NVMe, max 92.16 TB  
        \item  Up to 8 x 2.5-inch SAS/SATA/NVMe, max 122.88 TB 
    \end{itemize}
    \item Power supply: 
    \begin{itemize}
        \item 3200 W Titanium 277–305 VAC or 336 HVDC, hot swap redundant
        \item  2800 W Titanium 200–240 VAC 
        or 240 HVDC, hot swap redundant
        \item  2400 W Platinum 100–240 VAC or 240 HVDC, hot swap redundant 
    \end{itemize}
    \item External storage requirement
    \begin{itemize}
        \item Considering the same storage to compute capability ratio as in Microsoft inference datacenters ~\cite{microsoft2024nccadsh100v5}: 404 IOPS/TFLOPS 
        \item Taking example of Storage SuperServer SSG-121E-NE3X12R nodes (8 x 6.4TB SSD) 
        \item 1 x 42U storage rack = 42 x 8x900,000IOPS =  302,400,000IOPS

        \item 1 storage rack can serve up to:  302,400,000IOPS / (404 IOPS/TFLOPS ) =  6 x 140.4PFLOPS racks 
        \item 1 storage rack serves 6 compute racks.   
    \end{itemize} 
    \item Peak power per rack: \textbf{58.8kW/rack} for a 42U rack filled with PowerEdge R760xa nodes.
     \item  Fans: 
    \begin{itemize}
        \item Standard (STD) fan  
        \item  Up to six hot plug fans 
    \end{itemize}
    \item Available from July 2024 
\end{itemize}

\vspace{0.5cm} \noindent \textbf{IBM’s Gen AI~\cite{gershon2024infrastructure}}

\begin{itemize}
    \item Estimated compute capability: 30PFLOPS per rack  
    \item 6 servers per rack 
    \begin{itemize}
        \item 8 x 80GB A100 GPUs per node 
        \item 4,992TFLOPS per node 
        \item Connected to each other by NVLink and NVSwitch 
        \item  CPU: 3rd Generation Intel Xeon Scalable processors (Ice Lake)
        \item Memory: 1.5TB of DRAM
        \item Storage: 4x 3.2TB NVMe drives
    \end{itemize}
    \item External storage requirements:  
    \begin{itemize}
       \item Considering the same storage to compute capability ratio as in Microsoft inference datacenters ~\cite{microsoft2024nccadsh100v5}: 404 IOPS/TFLOPS 
        \item Taking example of Storage SuperServer SSG-121E-NE3X12R nodes (8 x 6.4TB SSD) 
        \item 1 x 42U storage rack = 42 x 8x900,000IOPS =  302,400,000IOPS

        \item 1 storage rack can serve up to:  302,400,000IOPS / (404 IOPS/TFLOPS ) =  25 x 30PFLOPS racks 
        \item 1 storage rack serves 25 compute racks.   
    \end{itemize} 
    \item Power demand: \textbf{40kW/rack }
    \item  Generation : 2023 
\end{itemize}

\vspace{0.5cm} \noindent \textbf{Google GDC Edge Rack (with GPUs)~\cite{CDGHardware}}
\begin{itemize}
    \item AI/ML  (INFERENCE) or graphic-intensive workloads 
    \item 1 rack = 6 GPU-enabled servers
    \begin{itemize}
        \item 96 vCPUs (16 core CPUs) per node
        \item Dual NVIDIA Tesla T4 GPUs per node
        \item Memory = 256GB per node 
        \item 4 TB SSD per node.
    \end{itemize}
    \item Rack dimensions: 80 inches (height) x 48 inches (dept) x 24 inches (Width). We estimate it to be a Around 46U (1U is 1.75” height) 
    \item 2 Top of Rack switches (10 Gigabit and 1 Gigabit Ethernet ports) 
    \item Power draw: \textbf{16.5kW/rack}
    \item Cooling technic: Air cooled  
    \begin{itemize}
        \item Temperature between 15°C and 31
        \item Relative humidity between 30\% and 70\% 
    \end{itemize} 
    \item Introduced on October 12, 2021 
\end{itemize}

\vspace{0.5cm} \noindent \textbf{ChatGPT inference configuration~\cite{ChatGPTInference}}
\begin{itemize}
\item Estimated compute capability : 18.1 EFLOPS (5PFLOPS per node~\cite{InspurNodePerf})
    \item 28,936 GPUs 
    \item 3,617 HGX A100   4U servers (Inspur NF5488A5 8x NVIDIA A100 HGX nodes~\cite{InspurNF5488A5}). Each node contains
    \begin{itemize}
        \item 2* AMD R OME Zen2 CPU 
        \item 8 GPUs per node 
        \item NVLink CPU-GPU connect
    \end{itemize}
    \item 40U racks. That is total 362 racks 
    \item External storage requirements:  
    \begin{itemize}
        \item Considering the same storage to compute capability ratio as in Microsoft inference datacenters ~\cite{microsoft2024nccadsh100v5}: 404 IOPS/TFLOPS 
        \item Taking example of Storage SuperServer SSG-121E-NE3X12R nodes (8 x 6.4TB SSD) 
        \item 1 x 40U storage rack = 40 x 8x900,000IOPS =  288,000,000IOPS

        \item Storage required by 362 compute racks:  
        18.1EFLOPS  x 404 (IOPS/TFLOPS) / 288,000,000IOPS  =  26 racks 
        \item 26 storage rack serves 363 compute racks.
    \end{itemize} 
    \item Power usage 
    \begin{itemize}
        \item 5.2kW TDP per node~\cite{InspurNF5488A5}
        \item For a 40U rack, that is  5.2kW x 10 = \textbf{52kW/rack }
    \end{itemize}

\item ChatGPT was launched in November 2022.
\end{itemize}

\vspace{0.5cm} \noindent \textbf{ Lenovo ThinkSystem SR670 V2 nodes~\cite{ThinkSystemSR670V2}}
\begin{itemize}
    \item Estimated compute capability: 2.15 PFLOPS per 42U rack OF 14 nodes~\cite{ThinkSystemSR670V2}
    \item 3U Rackmount servers. That is 14 nodes per 42U rack 
    \item GPUs (up to eight double-width GPUs with NVLink Bridge)
    \begin{itemize}
        \item Up to 4x double-wide, full-height, full-length; FHFL GPUs
        \item NVIDIA HGX™ A100 4-GPU with 4x NVLink connected SXM4 GPUs
    \end{itemize} 
    \item  CPUs: 2x 3rd Gen Intel® Xeon® Scalable processors per node 
    \item  Memory: up to 4TB using 32x 128GB 3200MHz TruDDR4 3DS RDIMMs  
    \item  Local storage: Up to 8x 2.5” Hot Swap NVMe SSDs
    \item External storage requirements:  
    \begin{itemize}
       \item Considering the same storage to compute capability ratio as in Microsoft inference datacenters ~\cite{microsoft2024nccadsh100v5}: 404 IOPS/TFLOPS 
        \item Taking example of Storage SuperServer SSG-121E-NE3X12R nodes (8 x 6.4TB SSD) 
        \item 1 x 42U storage rack = 42 x 8x900,000IOPS =  302,400,000IOPS

        \item 1 storage rack can serve up to:  302,400,000IOPS / (404 IOPS/TFLOPS ) =  348 x 2.15PFLOPS racks 
        \item 1 storage rack serves 348 compute racks.  
    \end{itemize}
    \item Power supply: 2.4kW per node   or\textbf{ 33.6kW/rack} (42U racks)
    \item Datasheet available from April 2021  	 
\end{itemize}

\vspace{0.5cm} \noindent \textbf{ Dell PowerEdge XE7745 nodes~\cite{DellPowerEdgeXE7745}} 
\begin{itemize}
    \item Estimated compute capability: 17.5 PFLOPS F16 per 42U rack (1,671 TFLOPS per node~\cite{NVIDIAGPUsPerf})
    \item 4U rack server 
    \item GPUs: 2 options 
    \begin{itemize}
        \item  8x PCIe Gen 5 x16 DW-FHFL up to 600W  
        \item 16x PCIe Gen 5 x16 SW-FHFL up to 75W 
    \end{itemize}
    \item CPUs: 2x 5th Generation AMD EPYC 9005 Series processors with up to 192 cores per processor 
    \item Memory:  24 DDR5 DIMM slots, supports RDIMM 2.3 TB max, speeds up to 6000 MT/s 
    \item Local storage: up to 8 x EDSFF E3.S Gen5 NVMe (SSD) max 122.88 TB 
    \item External storage requirements:  
    \begin{itemize}
       \item Considering the same storage to compute capability ratio as in Microsoft inference datacenters ~\cite{microsoft2024nccadsh100v5}: 404 IOPS/TFLOPS 
        \item Taking example of Storage SuperServer SSG-121E-NE3X12R nodes (8 x 6.4TB SSD) 
        \item 1 x 42U storage rack = 42 x 8x900,000IOPS =  302,400,000IOPS

        \item 1 storage rack can serve up to:  302,400,000IOPS / (404 IOPS/TFLOPS ) =  43 x 17.5PFLOPS racks 
        \item 1 storage rack serves 43 compute racks.  
    \end{itemize}
    \item Power supply
    \begin{itemize}
        \item 3200W Titanium 200-240 V AC or 240 V DC, hot swap redundant, per node
        \item That is \textbf{33.6kW/rack} (42U racks)
    \end{itemize}
    \item Available from January 2025 
\end{itemize}

\subsubsection{IT Hardware  Configurations in Cloud Datacenters}
\label{Sec:Cloud-SURVEY}
In this section, we describe reference datacenter configurations used in Cloud-type datacenters. Table~\ref{tab:AIDCConfig2} summarizes the key parameters of these configurations.

\renewcommand{\arraystretch}{2}
\begin{table}[H]
    \centering
    \resizebox{\textwidth}{!}{%
    \begin{tabular}{|c|c|c|c|c|c|}
    \hline
     \multirow{2}{*}{\textbf{Datacenter}} & \multirow{2}{*}{\makecell{\textbf{Rack size} \\ \textbf{(RU)}}}  & \multicolumn{2}{c|}{\textbf{CPU racks}} & \multicolumn{2}{c|}{\textbf{ Storage racks}}  \\
     \cline{3-6}
     &  & \textbf{\#racks} & \makecell{ \textbf{Peak power}\\ \textbf{(kW/rack)}} 
     & \textbf{\#racks} & \makecell{ \textbf{Peak power}\\ \textbf{(kW/rack)}}\\
     \hline 
     \rowcolor{gray!30}  
     Microsoft GreenSKU (2024)~\cite{10609689} & 42U &10 & 9 & 1 & 18.4 \\ \hline 

    \rowcolor{gray!30}  
    Microsoft Azure Stack HCI (2024)~\cite{McftHCI} & 42U & 5 & 17.6 & 1 & 18.4 \\ \hline 
    
    \rowcolor{gray!30}  
    Dell PowerEdge R660xs rack (2024)~\cite{r660xsSpec} & 42U & 1 & 29.4 & \multicolumn{2}{c|}{Local storage} \\ \hline 
        
    HP POD DC 44 (2022)~\cite{PODDC44} & 50U & 44 & 23 & 7 & 18.4 \\ \hline 

    Google distributed Cloud (2021)~\cite{CDGHardware} & 46U & 19 & 5 & 1 & 20.1 \\ \hline 
    
    Supercomputer Fugaku (2021)~\cite{Fujitsu2025FugakuSpecs} & 42U HPC & 414 & 72.2 & 11 & 13.8 \\ \hline 

    \end{tabular}
    }
    \caption{Summary of hardware configurations used in Coud datacenters. The grey cells show the most recent configurations (2024)}
    \label{tab:AIDCConfig2}
\end{table}

In the following, we show details of each reference datacenter configuration.

\vspace{0.5cm} 
\noindent \textbf{Microsoft GreenSKU~\cite{10609689}}

\begin{itemize}
\item 1 x 42U rack
\begin{itemize}
    \item 2U nodes
    \item CPU: AMD’s 128- core/256-thread x86 Bergamo \item Reused SSDs, and memory 
\end{itemize}
\item External storage estimation:
\begin{itemize}
      \item Taking example of the Storage SuperServer SSG-121E-NE3X12R nodes (8 x 6.4 SSD)~\cite{SuperServerSSG121ENE3X12R,SSD_datasheet}
        \item We consider a reduced storage node power of 438W corresponding to one full CPU (270W of CPU + 8x21W SSD). Storage TDP per racks = \textbf{18.4kW/rack} for a 42U rack
        \item Storage represents 18\% of total electrical power~\cite{masterVarshaRao} (see Table  4.2, page 23)
        \item Compute racks total power demand = 82\% x18.4kW /18\% = 83.8kW 
        \item  Number of compute racks:83.8/9kW = 10 racks
\end{itemize}
\item Peak power
\begin{itemize}
    \item Compute racks: 21 x 403 + 500 = 8963W,  
  \textbf{9kW/rack}
  \item Storage racks: \textbf{18.4kW/rack}
\end{itemize}
\item Generation: 2024
\end{itemize}

\vspace{0.5cm} \noindent \textbf{Microsoft Azure Stack HCI~\cite{McftHCI}}

\begin{itemize}
    \item 1x42U rack  
    \begin{itemize}
        \item 2-16 nodes per rack (e.g., Azure MC-760~\cite{Azure-MC-760} ) 
        \item CPU: 2x Intel Xeon Gold 6430 2.1 GHz, 64 cores, 128Threads 
        \item Memory 128GB
        \item Local Storage: 4x 800 GB SSD SAS = 3.2TB/node  
    \end{itemize}
    \item External storage estimation
        \begin{itemize}
            \item Taking example of the 438WW Storage SuperServer SSG-121E-NE3X12R nodes (8 x 6.4 SSD)~\cite{SuperServerSSG121ENE3X12R,SSD_datasheet}, storage TDP= \textbf{18.4kW/rack }for a 42U rack
        \item Storage represents 18\% of total electrical power~\cite{masterVarshaRao} (see Table  4.2, page 23)
        \item Compute racks total electrical power= 82\% x 18.4kW /18\% = 83.8kW 
        \item  Number of compute racks:83.8/17.6kW = 5 racks
        \end{itemize}
    \item Peak power demand
    \begin{itemize}
        \item Per compute node: 1.1kW 
        \item Per compute rack: 16*1.1kW = \textbf{17.6kW/rack} 
        \item Per storage rack : \textbf{18.4kW/rack}
    \end{itemize}
    \item Generation 2024
\end{itemize}

\vspace{0.5cm} \noindent \textbf{Dell PowerEdge R660xs nodes}

\begin{itemize}
\item  1U Rackmount node (A 42U rack can contain 42) nodes 
\item CPUs: Up to two 5th Generation Intel Xeon Scalable processor with up to 28 cores and 4th Generation Intel Xeon Scalable processor with up to 32 cores per processor 

\item  MEMORY: 16 DDR5 DIMM slots, supports RDIMM 1.5 TB max 

\item local storage: Up to 8 x 2.5-inch SAS/SATA/NVMe (HDD/SSD) max 122.88 TB 

\item Power supply  
\begin{itemize}
    \item 1800 W Titanium 200—240 VAC or 240 VDC  
    \item 1400 W Titanium 100—240 VAC or 240 VDC  
    \item 1400 W Platinum 100—240 VAC or 240 VDC  \item 1400 W Titanium 277 VAC or HVDC (HVDC stands for HighVoltage DC, with 336V DC)  
    \item 1100 W Titanium 100—240 VAC or 240 VDC 
    \item 1100 W -(48V — 60V) DC  
    \item 800 W Platinum 100—240 VAC or 240 VDC  
    \item 700 W Titanium 200—240 VAC or 240 VDC  
    \item 600 W Platinum 100—240 VAC or 240 VDC 
\end{itemize} 
\item Peak power per (42U) rack: 700kW x 42 = \textbf{29.4kW/rack}
\item Released in 2024 
\end{itemize}

\vspace{0.5cm} \noindent \textbf{Google distributed Cloud - Edge Rack (without GPUs)~\cite{CDGHardware}}

\begin{itemize}
    \item Large-scale general-purpose computing 
    \item 1 rack = 6 non-GPU servers  
    \item Rack dimensions: 80 inches (height) x 48 inches (dept) x 24 inches (Width) . We estimate the the racks around 46U (1U is 1.75” height) 
    Each server contains:
     \begin{itemize}
         \item  96vCPUs 
        \item RAM: 256 GB 
        \item 4 TB SSD
     \end{itemize}
    \item 2 Top of Rack switches (10 Gigabit and 1 Gigabit Ethernet ports) 
    
\item  Power draw  
\begin{itemize}
    \item Typical node power draw: 800W 
    \item That is \textbf{ 5kW/rack}
\end{itemize}

\item Central storage estimation:   
    \begin{itemize}
    \item Taking example of the 438W Storage SuperServer SSG-121E-NE3X12R nodes (8 x 6.4 SSD)~\cite{SuperServerSSG121ENE3X12R,SSD_datasheet}, storage TDP= \textbf{20.1kW/rack }for a 46U rack
        \item Storage represents 18\% of total electrical power~\cite{masterVarshaRao} (see Table  4.2, page 23)
        \item Compute racks total electrical power= 82\% *20.1kW /18\% = 91.6kW 
        \item  Number of compute racks:91.6/5kW = 19 x 5kW racks
    \end{itemize}
\item Introduced on October 12, 2021  
\end{itemize}

\vspace{0.5cm} \noindent \textbf{Supercomputer Fugaku~\cite{Fujitsu2025FugakuSpecs}}
\begin{itemize}
    \item 442.01 PFlop/s 
    \item Made of 158,976 CPU-powered nodes powered by  Armv8.2-A SVE 512 bit CPUs. Each node is made of: 
    \begin{itemize}
        \item 1 CPU of 48 cores + 2 assistant cores 
        \item HBM2 32 GiB, 1024 GB/s 
    \end{itemize}
    \item 2 nodes = 1 CPU Memory Unit (CMU) 
    \item 8 CMUs  = 1 brunch of blade (BoB)      
    \item 3 BoBs = 1 shelf  
    \item 8 shelves = 1 rack  
    \item 384 nodes per rack (or toal 414 racks)
    \item Liquid cooled with cold plates on the CMUs  
    \item Storage estimation
    \begin{itemize}
        \item 150PB shared storage based on  E1000 SSU-D device~\cite{FugakuStorage,FrontierStoragePower} 
        \begin{itemize}
            \item 1 E1000 SSU-D node = 106 x 16TB, or 1696TB per node
            \item 150PB/1.696 = 89 nodes 
            \item 1 rack = 8 nodes, hence 11 racks 
        \end{itemize}         
    \end{itemize}
    
    \item Power demand: 29.9 MW 
    \begin{itemize}
        \item 36.1 kW per half rack or \textbf{72.2kW/rack }~\cite{FugakuPowerConso}
        \item 1.728kW  per E1000 SSU-D device~\cite{FrontierStoragePower}
        \item Storage racks:  \textbf{13.8kW/rack }
    \end{itemize}
    \item Completed on March 9, 2021 
\end{itemize}

\vspace{0.5cm} \noindent \textbf{ HP Performance Optimized Datacenter (POD DC 44)~\cite{PODDC44}}
\begin{itemize}
    \item 44 x 50U racks 
    \item Power draw
    \begin{itemize}
        \item Compute racks: \textbf{23kW/rack}
    \end{itemize}
    \item Storage estimation  
    \begin{itemize}
        \item Taking example of the  438W Storage SuperServer SSG-121E-NE3X12R nodes (8 x 6.4 SSD)~\cite{SuperServerSSG121ENE3X12R,SSD_datasheet}
        item Storage represents 18\% of total electrical power~\cite{masterVarshaRao} (see Table  4.2, page 23)
        \item Storage total power: 44racks x 23kW x 12\% /82\% = 148.1kW = 338 nodes (438W each) =  7 x 50Uracks 
        \item Storage racks power:\textbf{29.6kW/rack }
    \end{itemize}
    \item Available from 2022 (based on the manual)
\end{itemize}

\subsection{Cooling System Hardware} 

Datacenter cooling systems are essential for keeping IT hardware at an optimal operation temperature, which prevents malfunctions and helps equipment last longer. These systems use a lot of energy – around 40\% of a datacenter's total power, as stated by Aljbour~\textit{et al.}~\cite{aljbour2024powering}. This is a significant amount, as the IT equipment itself uses about 40-50\% of the total energy. Therefore, continuous effort is required to improve cooling electrical efficiency. 

There are several types of datacenter cooling systems, mainly based on the nature of coolant used. Here's a breakdown of the main types:
\begin{itemize}
    \item \textbf{Air-based cooling systems}: these systems use air as cooling medium in the datacenter floor. They typically function by drawing cooler air from conventional fans or ventilation systems (computer room air conditioners, computer room air handlers...) and directing it over hot components (such as server racks). However, the heat capacity of air is low, making air cooling inefficient or non-applicable in high density datacenters.
    
    \item \textbf{Liquid-based cooling systems}: these systems use liquid (typically water) as cooling medium. The most popular method uses cold plates  -- usually aluminum or copper thermal interfaces -- placed on high-power components such as CPUs and GPUs. These plates have internal channels that allow cold water to circulate, absorbing heat from the processors and carrying it away. Liquid cooling allows to handle more heat, since water presents a higher  heat capacity. 
\end{itemize}

In practice, one may employ both air and liquid cooling. Thus, the heat generated in low and medium power components such as NICs, RAM... is removed by circulating cold air (e.g., from fans). 

The heat extracted from the IT hardware is often managed through a series of mechanical processes. Chillers and cooling towers or dry coolers work in concert, adapting their operation with regard to ambient weather, to ultimately release the heat into the environment. { 
Cooling towers and dry coolers are particularly sensitive to external temperature and humidity, experiencing reduced cooling capacity during hot weather compared to cooler periods. Consequently, datacenter cooling systems are typically designed to handle peak demand during the hottest times (daytime and summer). This results in excess capacity (over-sizing) for much of the day and throughout the year. Furthermore, the energy consumption of cooling equipment fluctuates with external conditions; for example, chillers can consume more than twice as much power during hot periods as they do in cooler weather periods.

In this document, we review liquid-based cooling systems. The cooling system is composed of \textit{Coolant Distribution Units }(CDUs)~\cite{VertivCoolChipCDU,VertivCoolPhaseCDU,VertivCoolChip600,VertivCoolChip100kW,VertivMegaModCDU}, \textit{chillers}~\cite{LiebertHPCW,VertivLiebertHPCW,VertivLiebertOFC} and a \textit{heat sink}. The most adopted technologies of heat sink are \textit{dry coolers}~\cite{ReftecoDryCooler,VertivLiebertLVC,VertivLiebertDrycooler} (water-efficient) and  \textit{evaporative cooling  towers} (energy efficiency) such as in~\cite{Evapco167B,EvapcoATUTUSS}. DCGen accommodates a variety of cooling equipment  options 
that differ in vendor, capacity, efficiency, and physical footprint. For a target scale, specific optimization objectives (e.g., power, space,...) guide the selection of the most suitable configuration. 

 
\subsection{Power Distribution Hardware}
A datacenter's power system provides continuous, stable and reliable power to the IT and cooling infrastructure. It is \added{typically} a 3-level  hierarchy~\cite{vertivAI2025} as shown in Figure~\ref{fig:Datacenter_design}.
Power enters first at a utility substation which transforms high voltage (typically over 110
kV~\cite{barroso2019datacenter}) to medium voltage (less than 50 kV~\cite{barroso2019datacenter}). Next, it is converted 
by  Main SwitchBoards (MSBs)~\cite{VertivPowerBoardUL891,VertivPowerBoardLV}, below 1kV~\cite{barroso2019datacenter}. The MSBs also receive input from \textit{diesel/gas generators} (backup power supply)~\cite{CatDieselGen,CatGasGen,Wartsila10MW,Guangling10MW,CatC18Diesel,MeccagenGenerators} during outages. Power enters the DC building via low-voltage lines  connected to  Uninterruptible Power Supplies (UPS) systems\cite{VertivLiebertAPM,VertivPowerUPS9000,VertivTrinergyUPS,VertivLiebertEXLS1,VertivPowerModuleH2}. It is then distributed at datacenter HVDC distribution (but 100-480V, much lower than grid).   On the DC floor, power goes to \textit{Power Distribution Units} (PDUs)~\cite{VertivLiebertSPM,VertivLiebertAPT,VertivLiebertTFX,VertivLiebertSTS2}, the
last layer in the transformation and distribution architecture above the power supplies that support each IT component.
DCGen accommodates a variety of power distribution equipment  options  that differ in vendor, capacity, efficiency, and physical footprint. For a target scale, specific optimization objectives (e.g., power, space,...) guide the selection of the most suitable configuration.

Downtime can lead to significant financial and operational losses, so power distribution systems are built with redundancy, scalability and fault tolerance in mind. 
It is also worth noting that every stage in the power conversion process entails power losses (up 15\% losses~\cite{barroso2019datacenter}). The electrical system is therefore not a trivial mechanism, \deleted{constantly aiming for improvement} \added{and needs constant improvements to ensure continuous and reliable power}.

\subsection{Summary}

To automate hardware configuration across diverse datacenter designs, we created DCGen, a tool that integrates various mathematical models. This tool can generate configurations \added{optimized for space or power, } based on two primary targets: (1) a specific number of racks (representing the desired compute capability), and (2) a target electrical power use. By inputting one of these targets, the tool produces a configuration that matches the specified requirement. 
\added{The output configuration includes IT hardware, cooling and power distribution infrastructures as shown in Figure~\ref{fig:DCGenFeatures}.} 
Furthermore, the tool produces  metrics such as datacenter power density, electrical power, and space requirements for that configuration.   In the following, Section~\ref{Sec:Modeling} elaborates on datacenter IT  -- both with compute capability target and target electrical power use--, and non-IT (cooling system and power distribution system) modeling.

\begin{figure}[hpbt]
    \centering
\includegraphics[width=\linewidth]{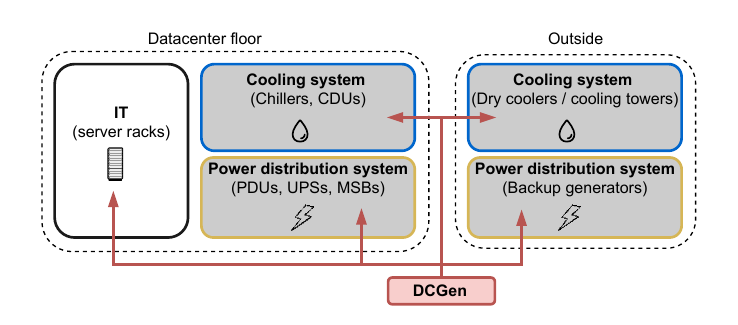}
    \caption{DCGen generates datacenter IT configuration, the cooling system and power distribution system for both IT load and the cooling system. }
    \label{fig:DCGenFeatures}
\end{figure}

\section{Modeling}
\label{Sec:Modeling}

\subsection{System Architecture}

Datacenters contain rows of server racks, supported by cooling systems for heat dissipation and a power distribution hierarchy ~\cite{stojkovic2025tapas,zhang2021flex,vertivAI2025,alkrush2024data,barroso2019datacenter}. Figure~\ref{fig:Datacenter_design} illustrates the design implemented in DCGen \deleted{V1.0} \added{V1.1}.  It is an adaptation of the RD027 architecture in~\cite{vertivAI2025},  with the types and power density of racks changing with the datacenter type and reference design.

\begin{figure}[hbpt]
    \centering
    \includegraphics[width=\linewidth]{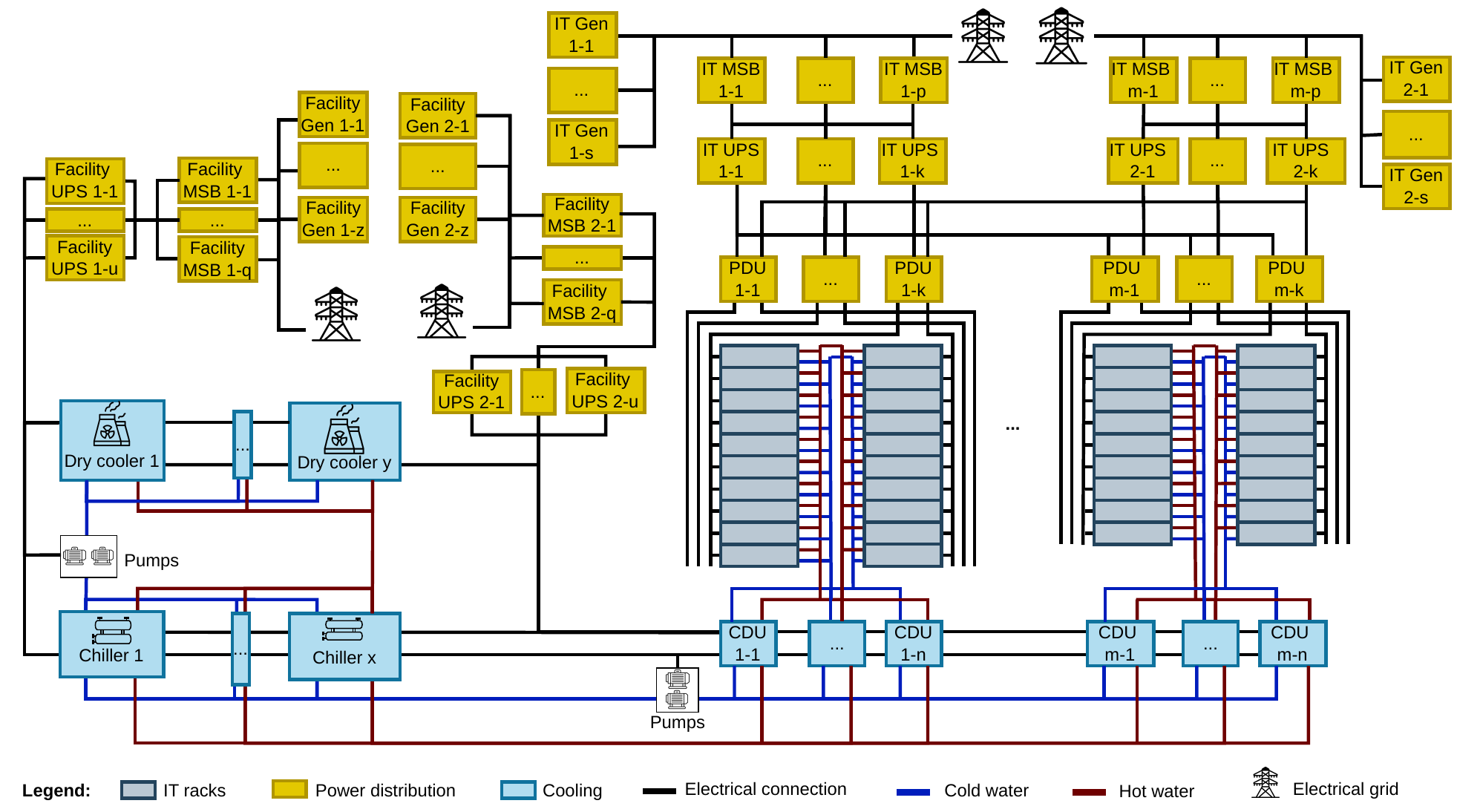}
    \caption{Datacenter design including IT, cooling and power system configurations.The architecture is based on~\cite{vertivAI2025}.}
    \label{fig:Datacenter_design}
\end{figure}

\subsection{Canonical IT Hardware Models}
\label{sec:CANONICAL-IT-MODEL}
\deleted{To enable a global view study of datacenter configurations, we aggregated several reference hardware configurations (see Section~\ref{SEC::BASE-DATACENTERS})} \added{We aggregated several reference IT configurations (see Section~\ref{SEC::BASE-DATACENTERS}) to build datacenters that capture characteristics from multiple hardware setups}. We consider three configurations: \textit{today's datacenter hardware configurations} that are the aggregation of recent configurations (2024), 2027 and 2029 datacenter configurations projecting datacenter designs for 2027 and 2029 respectively. All datacenters utilize 42U racks. 

At the rack level, the peak electrical power and the number of racks of each node type are summarized in Table~\ref{tab:CaseStudy}.

\renewcommand{\arraystretch}{1.5}
\begin{table}[hbpt]
    \centering
    \resizebox{\linewidth}{!}{%
    \begin{tabular}{|c|c|c|c|}
    \hline
    \textbf{Type} & \textbf{Today's datacenters} & \textbf{2027 datacenters}  & \textbf{2029 datacenters}\\
    \hline
    \hline
    \multicolumn{4}{|c|}{\textbf{datacenter level configurations}}  \\
    \hline
         AI training datacenter & \makecell{GPU racks: 300 \\ Storage racks: 38 } & \makecell{GPU racks: 300 \\ Storage racks: 265 } & \makecell{GPU racks: 300 \\ Storage racks: 373 } \\
    \hline
    AI inference datacenter & \makecell{CPU-GPU racks: 100 \\ CPU racks: 100 \\ Storage racks: 23} & \makecell{CPU-GPU racks: 100 \\ CPU racks: 100 \\ Storage racks: 53 }  & \makecell{CPU-GPU racks: 100 \\ CPU racks: 100 \\ Storage racks: 146 } \\
    \hline
    \makecell{Mixed AI training \\ and inference datacenter} & \makecell{GPU racks: 100 \\ CPU-GPU racks : 100 \\ Storage racks: 22 } & \makecell{GPU racks: 100 \\ CPU-GPU racks: 100  \\ Storage racks: 104  }  & \makecell{GPU racks: 100 \\ CPU-GPU racks: 100  \\ Storage racks: 68  }\\
    \hline
    Cloud datacenter & \makecell{CPU racks: 100 \\ Storage racks: 22 } & \makecell{CPU racks: 100 \\ Storage racks: 60 } & \makecell{CPU racks: 100 \\ Storage racks: 60 } \\
    \hline 
    \hline
    \multicolumn{4}{|c|}{\textbf{Racks peak electrical power}}  \\ \hline
    GPU racks      & 158kW/rack  &  600kW/rack  &  1000kW/rack \\ \hline
    CPU-GPU racks  & 50kW/rack   &  90kW/rack  & 152.1kW/rack \\ \hline
    CPU racks      & 18.7kW/rack &  50kW/rack  & 59.4kW/rack \\ \hline
    Storage racks  & \multicolumn{2}{|c|}{ \makecell{29.7kW/rack in AI datacenters \\ 18.4W/rack in Cloud datacenters}}  &   \makecell{35.3kW/rack in AI datacenters \\ 21.9W/rack in Cloud datacenters} \\ \hline
    \end{tabular}
    }  
    \caption{Datacenter \deleted{canonical} \added{IT hardware} configurations.}
    \label{tab:CaseStudy}
\end{table}

In the following, we give a breakdown of the canonical hardware configurations used in DCGen.
\begin{enumerate}
\item Racks of \textbf{GPU nodes} 
    \begin{itemize}
        \item \textbf{Today's datacenter}: it is the average of these datacenters: (1)NVIDIA DGX SuperPOD With DGX GB200 Systems, (2) xAI COLOSSUS, (3) El Capitan. The peak electrical power is estimated to be 158kW/rack. 
    
        \item \textbf{2027 datacenter}: our reference configuration is the NVIDIA Rubin Ultra NVL576 rack systems announced for 2027.  The peak electrical power is 600kW/rack.

        \item \textbf{2029 datacenter}: We base these datacenters on predictions of 1MW racks by 2029~\cite{ElongatedMusk2023,Vertiv2024AI}.
    \end{itemize}

\item Racks of \textbf{CPU-GPU nodes}
    \begin{itemize}    
        \item \textbf{Today's datacenter}:  it is the average of these datacenters (1)  IBM's Gen AI, (2)  Supermicro cloud-scale inference rack systems, (3) racks filled with Dell PowerEdge R760xa. The peak electrical power is estimated to be 50kW/rack. On average, a rack of such configuration delivers 75.3PFLOPS.

         \item \textbf{2027 datacenter}:  
        As stated in this report, \textit{AI workloads are pushing requirements into the 60 to 120 kW range}~\cite{MattVincentCoolinIT}. As it does not specify the actual share of inference (that uses CPU-GPU nodes), we consider the average number which is 90kW/rack and assume it to remain valid in 2027. To estimate future storage requirement, we assume that storage racks will account for the same proportion of total electrical power as in the today's configuration.

        \item \textbf{2029 datacenter}:  the International Energy Agency states that Electricity consumption in accelerated servers, which is mainly driven by AI technology adoption, is projected to grow by 30\% annually~\cite{iea2025energy}. Hence, we assume that power density grows by 30\% per year from 2027. 
        
    \end{itemize}
\item Racks of \textbf{CPU nodes}
    \begin{itemize}    
         \item \textbf{Today's datacenter}:
         it is the average of these datacenters: (1) Microsoft GreenSKU model,  (2) Microsoft Azure Stack HCI, (3) a 42U rack filled with Dell PowerEdge R660xs nodes. The peak electrical power is estimated to be 18.7kW/rack.

         \item \textbf{2027 datacenter}:
         As indicated in~\cite{DuffFutureOfDC,AmandaFutureOfDC}, hyper-scale datacenters peak power is  projected to reach 50 kW/rack by 2027. 

         \item \textbf{2029 datacenter}:
        the International Energy Agency states that Electricity consumption in conventional servers is projected to grow by 9\% annually~\cite{iea2025energy}. We assume 9\% annual growth from 2027, of CPU rack power density to project future configurations. 
    \end{itemize}
\item Racks of \textbf{Storage nodes}
    \begin{itemize}    
        \item \textbf{Today's datacenter}: We consider a 42U rack filled with Supermicro SuperServer SSG-121E-NE3X12R nodes. The peak electrical power is estimated to be 18.4kW/rack.
        
        \item \textbf{2027 datacenter}: We consider a 42U rack filled with Supermicro SuperServer SSG-121E-NE3X12R nodes. The peak electrical power is estimated to be 18.4kW/rack.
        
        \item \textbf{2029 datacenter}: We assume 9\%  annual growth from 2027, as CPU racks. 

    \end{itemize}
\end{enumerate}

At the datacenter level, we made some assumptions in AI inference, and mix AI training and inference configuration
\begin{itemize}
    \item \textbf{AI inference}: we assume that the datacenters have as many CPU-GPU racks as CPU racks, and estimate the amount of storage racks required. 
    \item \textbf{Mixed AI training and inference}: we assume that the datacenters have as many GPU racks as CPU-GPU racks, and estimate the amount of storage racks required. 
\end{itemize}

\subsection{DCGen Model Parameters}
\subsubsection{DCGen Inputs}
Users input the following parameters:
\begin{itemize}
    \item Datacenter target: it can be one of the following two: 
    \begin{itemize}
        \item     The total number of racks in the target datacenter that represents compute capability 
        \item Target power capacity (MW)
    \end{itemize}
    \item The datacenter type:  AI training, Mixed AI training and inference, AI inference, Cloud.
    \item The target year of operation: 2024, 2027, 2029
    \item Cooling and Power Distribution Systems redundancy
    \item Cooling and Power Distribution Systems safety margins
\end{itemize}

\subsubsection{DCGen Internal Parameters}
Internal parameters to DCGen are the following:
\begin{itemize}
    \item Reference representative IT hardware configurations. There are several options among real-world datacenter designs and canonical models resulting from averaging several of those systems. 

    \item In each reference design, the peak rack power demand is specified (kW/rack).

    \item The rack sizes (Rack Units) and average floor space utilization per rack (m$^2$)

    \item Cooling system (CDU, chillers, dry coolers, evaporative cooling towers), and Power Distribution System (PDUs, UPSs, MSBs, backup generators) hardware. Each type of hardware contains diverse options that may be suitable in different use cases. 

    \item Physical dimensions of each cooling and power system equipment, and access areas (m$^2$).
\end{itemize}

\subsubsection{DCGen Outputs}
DCGen outputs the following key parameters
\begin{itemize}
    \item The IT hardware configuration that meets design targets, i.e., the distribution of racks per node types.

    \item Datacenter power density, i.e., power utilization per unit of space (kW/m$^2$)

    \item Datacenter IT electrical power (MW)

    \item IT (\added{or White}) space utilization

    \item Cooling and power distribution systems hardware configuration, i.e, the number and specifications of the suitable equipments at each level of cooling and power hierarchy

    \item Cooling and power distribution power demand (MW)

    \item Gray space, i.e., space requirements of cooling and power systems\textsuperscript{\ref{footnote_space}}.
\end{itemize}

\subsection{ Modeling Datacenter IT driven by Compute Capability}
\label{SEC::DatacenterModelWithRackNbr}

In this section, we describe the modeling of datacenter configurations based on the reference hardware presented in Section~\ref{SEC::BASE-DATACENTERS}. The model enables the scaling of datacenters by employing rack-level specifications and target compute capability, thereby automating datacenters IT configuration. Moreover, this model facilitates the analysis of key metrics including power density, electrical power, and space requirements.
The inputs, model, and outputs  are represented in Figure~\ref{fig:DCGen}.

\begin{figure}[H]
    \centering
\includegraphics[width=\linewidth]{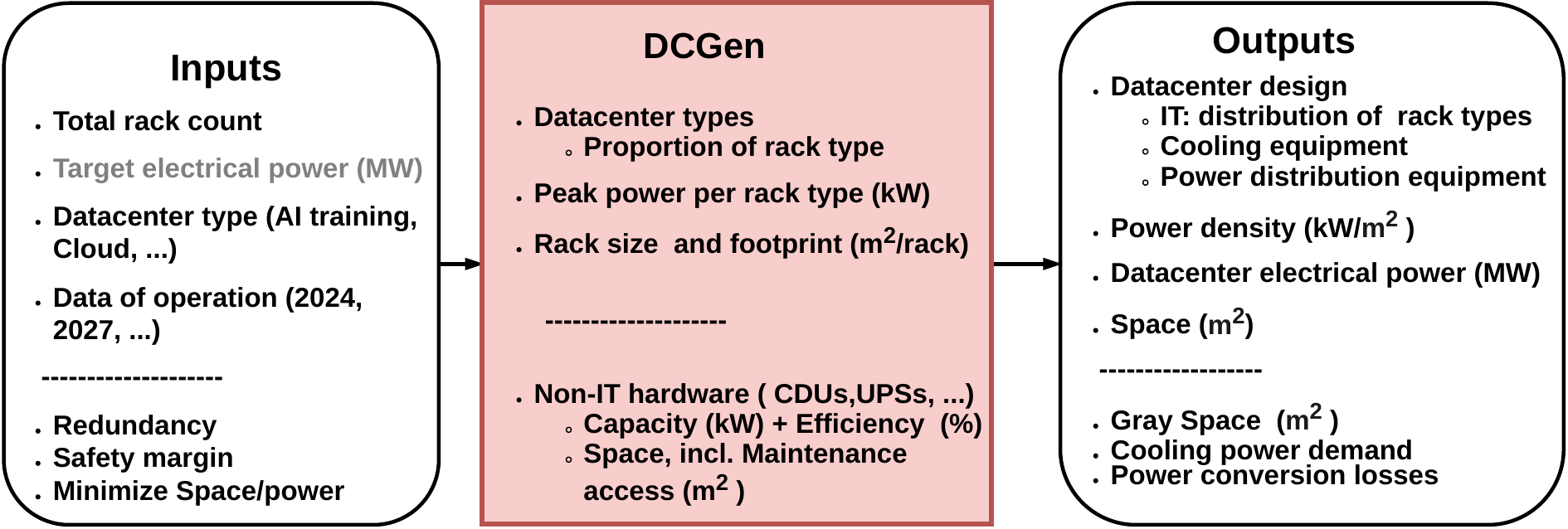}\\[-5pt]
    \caption{DCGen takes either a desired number of racks or a target electrical power as input. It then produces hardware configurations and other crucial metrics.} 
    \label{fig:DCGen}
\end{figure}

To formulate the DCGen model, let's define the following parameters: 

\begin{itemize}
    \item  $N_{rack}$: total number of racks in the target datacenter.
    \item datacenter type: AI training, AI inference, mixed AI training and inference, Cloud ...
    \item date-of-operation: 2024, 2027, \added{2029}
     \item $A_{rack}$: average floor space per rack ($m^2/rack$\deleted{ or $sqft/rack$}), in the datacenter.

\item $N_{0,rack,type1}$, $N_{0,rack,type2}$, ...: a breakdown of the number of racks per nodes type, in the reference configuration. For instance, in an AI training datacenter, type1 is GPU node racks and type2 stands for storage node racks. 
\item $P_{0,rack, type1}^{\max}$, $P_{0,rack, type2}^{\max}$, ...:  max electrical power per rack (kW) type in the reference configuration. 
    \item $RU_{0,rack}$: rack size specified in rack units (abbreviation U or RU). The rack unit size is based on a standard rack specification as defined in EIA-310. For example, a typical full-size rack cage is 42U high. However, some configurations may used 48U, 50U... This parameter is particularly important for comparing different datacenter configurations, by scaling them to the same rack size.  
\end{itemize}

DCGen uses these parameters, representing a reference datacenter's IT configuration, to generate new datacenters' setup. The reference configurations may either be derived from an existing system as detailed in Section~\ref{SEC::BASE-DATACENTERS}, or an aggregation of several systems into a canonical model.

\subsubsection{Storage Requirement }
\label{Sec::BASE-STORAGE-REQUIREMENT}
\paragraph{Storage for AI Training:}
in most of the reference IT configurations, the number of storage racks is not specified in the official documentation.
Hence, we estimated it using the total storage volume and a typical storage node. We consider the Supermicro 1U Storage SuperServer SSG-121E-NE3X12R nodes with the storage capacity of 8 x 6.4 SSD~\cite{SuperServerSSG121ENE3X12R,SSD_datasheet} as a reference. We estimated the peak electrical power of each storage node to be 708W, which corresponds to when the processors run at maximum capacity and the 8 SSD cards write with maximum speed. The model (Equation~\ref{Eqn::BaseStorageRacks}) divides the total volume of storage by the capacity of the reference node (which gives the total number of 1U storage nodes) and then divides the total nodes by the capacity of a rack.
For instance, a 42U rack filled with storage nodes contains 42 nodes.

\begin{equation}
\label{Eqn::BaseStorageRacks}
     N_{0,rack,storage} = \lceil \frac{\text{DC total storage}}{8 \times 6.4TB \times RU_{0,rack}}\rceil
\end{equation}

In the configurations where we total amount of storage is not specified in the official documentation, we consider the same proportion of GPU racks power/storage racks power as in xAI COLOSSUS~\cite{xAIColossus}. 
The xAI COLOSSUS datacenter contains around 500PB  of storage (estimated to 203 x 34kW racks) and 1563 x 100kW GPU racks (100,000 GPUs). Hence, \textbf{storage racks represent 4.2\% of the total electrical power}. Hence, we assume that 4.2\% of the datacenter electrical power comes from storage racks. The number of storage racks is given as in Equation~\ref{Eqn::StorageEstimation}.

\begin{equation}
\label{Eqn::StorageEstimation}
     N_{0,rack,storage} = \lceil \frac{4.2\% \times N_{0,rack,type1 } \times P_{0,rack, GPU}^{\max} }{95.8\% \times RU_{0,rack} \times 708W}\rceil
\end{equation}

\paragraph{Storage for AI Inference:}
to estimate the storage usage of AI inference infrastructures, we assume the same ratio between external storage IOPS (Input/Output Operations Per Second) and the datacenter compute capability (teraflops), as in Microsoft inference datacenters~\cite{microsoft2024nccadsh100v5}. Specifically, in the NCCads\_H100\_v5 sizes series, each Nvidia H100 GPU (94GB) of 1,979TFLOPS~\cite{NVIDIAGPUsPerf} can use up to 8x100,000 IOPS disks. That is a ratio of 404 IOPS/TFLOPS. 

Moreover, we consider the Supermicro 1U Storage SuperServer SSG-121E-NE3X12R nodes with the storage capacity of 8 x 6.4 SSD~\cite{SuperServerSSG121ENE3X12R,SSD_datasheet} as a reference. Each disk performs Random Write with 900,000 IOPS. 
Given an AI inference datacenter with the compute capability in TFLOPS, the number of storage racks required is presented in Equation~\ref{eqn::StroageIOPS}.

\begin{equation}
    \label{eqn::StroageIOPS}
    N_{0,rack,storage} = \frac{\text{DC compute capability (TFLOPS)} \times 404 \text{ (IOPS/TFLOPS)} }{ 8 \times 900,000 \text(IOPS) \times RU_{0,rack}  }
\end{equation}

The electrical power per rack can be estimated as in Equation~\ref{Eqn::StorageRackTDP}. 
\begin{equation}
    \label{Eqn::StorageRackTDP}
     P_{0,rack, storage} = RU_{0,rack} \times 708W
\end{equation}

 \paragraph{Storage for Cloud Datacenters}: 
In the configurations where the total storage capacity of a Cloud-type datacenter is known, the number and peak power demand of storage racks is estimated as for AI training (Equation~\ref{Eqn::StorageEstimation}).
 We consider the Supermicro 1U Storage SuperServer SSG-121E-NE3X12R nodes with the storage capacity of 8 x 6.4 SSD~\cite{SuperServerSSG121ENE3X12R,SSD_datasheet} as a reference. We estimated the peak electrical power of each storage node when used in Cloud datacenters to be 438W, which corresponds to one processor running at maximum capacity and the 8 SSD cards write with maximum speed.

However, when the storage capacity is not known, we estimate the number of storage racks based on the proportion of storage power demand in Cloud datacenters. 
According to empirical data from~\cite{masterVarshaRao} (see Table  4.2, page 23), storage represents 18\% of total electrical power in Cloud datacenters. The number of storage racks is estimated as in Equation~\ref{Eqn::StorageEstimationCloud}.

\begin{equation}
\label{Eqn::StorageEstimationCloud}
     N_{0,rack,storage} = \lceil \frac{18\% \times N_{0,rack,CPU } \times P_{0,rack, CPU}^{\max} }{82\% \times RU_{0,rack} \times 438W}\rceil
\end{equation}

The electrical power per rack can be estimated as: 
\begin{equation}
     P_{0,rack, storage} = RU_{0,rack} \times 438W
\end{equation}

\subsubsection{Model Outputs: IT Space, Power, and Density}



\paragraph{Type and Number of Racks}:
as described above, the model takes as input the total number of racks in the datacenter, and using the reference configuration, Equation~\ref{Eqn::NBRTYPE1} calculates the number of nodes and racks of each type.

\begin{subequations}
\begin{gather}
    \label{Eqn::NBRTYPE1}
        N_{rack,type \text{ }i} = \lceil \frac{N_{0,rack,type \text{ } i} }{ \sum_k N_{0,rack,type  \text{ }k} }  \rceil . N_{rack} \;   \text{ ; for k and i in node types}
\end{gather}
\end{subequations}

\paragraph{Electrical Power}:
the datacenter electrical power is estimated by adding the peak power demand of the racks as presented in Equation~\ref{Eqn::DCPowerCapacity}. The electrical power is converted in MW by (division by $10^3$). 

\begin{equation}
\label{Eqn::DCPowerCapacity}
     P_{DC}^{\max} = \sum_i ( N_{rack,type\text{ } i} \times \frac{P_{rack,type\text{ } i}^{\max}}{10^3} )  \; \text{ ; for i in node types}
\end{equation}
   
\paragraph{Power Density}:
we define power density as the amount of peak electricity usage per unit of space ($kW/m^2$\deleted{or $kW/sqft$}). This indirectly gives an insight on the amount of space required to build a certain datacenter. Power density is estimated as in Equation~\ref{Eqn::PoweerDensity}. 

\begin{equation}
\label{Eqn::PoweerDensity}
\begin{aligned}
       \text{Power density} &= \frac{ \sum_i N_{rack,type\text{ } i} \times P_{rack,type\text{ } i}^{\max} }{ A_{rack} \times N_{rack} }  \; \text{ ; for i in node types} \\[10pt]
        &= \frac{ \sum_i N_{0,rack,type\text{ } i} \times P_{rack,type\text{ } i}^{\max} }{ A_{rack} \times (\sum_i N_{0,rack,type\text{ } i}) } \; \text{ ; for i in node types} 
\end{aligned}
\end{equation}

\paragraph{IT Floor Space (or White space)}:
the datacenter floor space ($m^2$\deleted{or $sqft$}) can be calculated in two ways: by dividing the total electrical power by the power density, or by multiplying the number of racks by the space each rack occupies, as represented in Equation~\ref{Eqn::FloorSpace}

\begin{equation}
\label{Eqn::FloorSpace}
    \text{Floor space} = \frac{10^3 \times P_{DC}^{\max}}{\text{Power density}} = N_{rack} . A_{rack}
\end{equation}

\subsubsection{Comparing Datacenters with Different Rack Types (sizes)}
Some configurations use different rack sizes (e.g. HPC racks, 42, 40, 46, 48 RU racks). To establish  uniform rack size to compare different configurations, we estimate electrical power for a given rack type (typically 42U).

Let $RU_{rack}$ be the chosen rack. The electrical power per rack type is estimated as follows:

\begin{equation}
    P_{rack,type\text{ i}}^{\max} = \frac{RU_{rack}}{RU_{0,rack}} .  P_{0,rack,type\text{ i}}^{\max} \; \text{ ; for i in node types}
\end{equation}

For HPC datacenters (like Fugaku, El Capitan,...), we multiplied the peak power demand by 2/3.

\subsection{Modeling Datacenter IT Configuration driven by Target Electrical Power}
\label{SEC::DatacenterModelWithPower}

Datacenter design can be driven by a target of electrical power use rather than a specific compute capability (number of racks).
In this section, we describe the modeling of datacenter IT hardware configurations based on the reference presented in Section~\ref{SEC::BASE-DATACENTERS}, and a target electrical power use. In addition to automating hardware configuration to meet the power requirement, the model allows to analyze datacenters power density and space requirement.
The inputs, the model, and outputs are represented in Figure~\ref{fig:DCGen}.

In this section, we use the same notation as in Section~\ref{SEC::DatacenterModelWithRackNbr}, for IT modeling. The reference configurations may either be an existing system as detailed in Section~\ref{SEC::BASE-DATACENTERS},  or an aggregation of several hardware configurations into a canonical model.

In reference configurations involving data storage racks which are unspecified, we use the methodology described in Section~\ref{Sec::BASE-STORAGE-REQUIREMENT} to make estimation. 

The model outputs are: IT hardware and White Space requirements.

\paragraph{Power density}:
the datacenter power density ($kW/m^2$\deleted{ or $kW/sqft$}) is estimated as in Equation~\ref{Eqn::PoweerDensity2}, using the reference configuration racks distribution. 

\begin{equation}
        \label{Eqn::PoweerDensity2}
       \text{Power density} = \frac{ \sum_i N_{0,rack,type\text{ } i} \times P_{rack,type\text{ } i}^{\max}  }{A_{rack} \times ( \sum_i N_{0,rack,type\text{ } i}) }  \; \text{ ; for i in node types}
\end{equation}

\paragraph{IT space (or White space)}:
the datacenter floor space ($m^2$\deleted{ or $sqft$}) can be calculated by dividing the total electrical power by the power density Equation~\ref{Eqn::FloorSpace2}.

\begin{equation}
\label{Eqn::FloorSpace2}
    \text{Floor space} = \frac{ 10^3 \times P_{DC}^{\max}}{\text{Power density}} 
\end{equation}

\paragraph{Type and Number of Racks}:
The total number of racks required to design a datacenter with the target electrical power ($P_{DC}^{\max}$) is estimated as in Equation~\ref{Eqn::RacksNumber}. The model divides the total floor space of the datacenter by the space required by each rack.

\begin{equation}
\label{Eqn::RacksNumber}
    N_{rack} = \frac{\text{Floor space}}{A_{rack}} = \frac{ 10^3 \times P_{DC}^{\max}}{\text{Power density} \times A_{rack}} 
\end{equation}
 
Equation~\ref{Eqn::NBRTYPE2}  calculate the rack distribution for each node type. The models operate under the assumption that the percentage of each rack type remains constant between the base and target datacenters (e.g., if the reference datacenter contains 70\% of GPU racks, the target datacenter maintains 70\% of GPU racks). Hence, the hardware is scaled proportionally.  

\begin{subequations}
\begin{gather}
    \label{Eqn::NBRTYPE2}
        N_{rack,type\text{ } i} = \lceil \frac{  N_{0,rack,type\text{ } i} }{\sum_k N_{0,rack,type\text{ } k} }  \rceil . N_{rack} \; \text{ ; for k and i in node types}
\end{gather}
\end{subequations}

\subsection{Modeling Datacenter Cooling and Power Systems}
\label{Sec::NON-IT-CONFIGURATION}

\subsubsection{N+r Redundancy Scheme}
Datacenter cooling and power systems  design typically falls into two categories~\cite{vertiv360ai}: rack-level and datacenter-level hardware (See Figure~\ref{fig:Datacenter_design}). Components such as CDUs and PDUs are configured at the rack level. Racks are organized into pods, with each pod consisting of $N_{\text{row,pod}}$ rows. Let $N_{\text{rack,row}}$ denote the number of racks per row, and $P_{rack}^{\max}$ the peak power demand per rack. Considering an $N+r$ (e.g., N+1, N+2, etc.) redundancy  and a safety margin $sm$ (equipment oversizing to support extreme emergency scenarios or allow for future IT expansion). The number of hardware units per pod is determined using Equation~\ref{Eqn::Hardware_count_per_rack_group_N_plus_r}. The total hardware in the datacenter is then obtained by multiplying the number of units per pod by the total number of pods (Equation~\ref{Eqn::Total_hardware_count}), where $P_{DC}^{\max}$ represents the IT peak power.

\begin{subequations}
    \begin{gather}
\label{Eqn::Hardware_count_per_rack_group_N_plus_r}
        N_{l, pod} = \lceil\frac{ (1+sm). N_{\text{row,pod}}. N_{\text{rack,row}}. P_{rack}^{\max} }{P_{l}^{\max}}\rceil + r  \qquad  ; l \in \text{ \{CDU, PDU\}} \\
        \label{Eqn::Total_hardware_count}
        N_l = \lceil \frac{P_{DC}^{\max}}{P_{rack}^{\max} .N_{\text{rack,row}}. N_{\text{row,pod}} }  \rceil . N_{l, pod}
  \qquad  ; l \in \text{ \{CDU, PDU\}} 
    \end{gather}
\end{subequations}

At the datacenter level, the  number of cooling and power component required to support the IT load is determined using Equation~\ref{Eqn::Equipment_count_DC_level_IT}. Moreover, the cooling system introduces additional power demand, also handled by UPSs, MSBs, and backup generators (Equation~\ref{Eqn::Equipment_count_DC_level_Facility}).

\begin{subequations}
    \begin{gather}
    \label{Eqn::Equipment_count_DC_level_IT}
       N_{l,\text{IT}} = \lceil \frac{ (1+sm) . P_{DC}^{\max} }{P_l^{\max}} \rceil + r
      \qquad  ; l \in \text{ \{chiller, cooler, UPS, MSB, Gen\}} \\ 
\label{Eqn::Equipment_count_DC_level_Facility}
        \resizebox{0.4\linewidth}{!}{$N_{l, \text{Facility}} = \lceil \frac{ (1+sm). \sum_k P_{k}^{\max} }{P_l^{\max}} \rceil + r $}
  \quad  ; \resizebox{0.5\linewidth}{!}{$l \in \text{ \{UPS, MSB, Gen\}} , k \in  \{CDU, chiller, cooler\} $} \\
  \label{Eqn::gray_space}
  \text{Gray space}=  \sum_l (1+ \lambda_{l}) (N_{l,\text{IT}} +N_{l,\text{Facility}} ) . A_l    
    \end{gather}
\end{subequations}

Gray space is the area occupied by cooling and power infrastructures. It is estimated by summing the space footprint of each non-IT  component $l$ (Equation~\ref{Eqn::gray_space}). This footprint includes both the physical surface area of the unit ($A_l$), and additional space required for maintenance access area, which is assumed to scale with the equipment size (represented by the proportionality factor $\lambda_{l}$).

\subsubsection{xN/y Redundancy Scheme}
DCGen supports redundancy schemes of the xN/y type, such as 4N/3, 2N, etc. As shown by the authors of~\cite{Flex}, under an xN/y redundancy configuration, each hardware unit delivers only a fraction of its nominal capacity, which is quantified in Equation~\ref{Eqn::Power_limit}. For example, in a 2N setup, each piece of equipment provides only half of its rated capacity, effectively requiring the procurement of twice the amount of hardware needed to meet the datacenter's operational demand.

The number of CDUs and PDUs required per pod is determined using  Equation~\ref{Eqn::Hardware_count_per_rack_group_xN_over_y}, which calculates the ratio of the pod's peak power demand (including the safety margin) to the effective deliverable capacity per unit. The total number of hardware units needed to support the entire datacenter's IT load is then computed in Equation~\ref{Eqn::Total_hardware_count_xn_over_y}, by multiplying the hardware count per pod by the total number of pods in the datacenter.

\begin{subequations}
    \begin{gather}
    \label{Eqn::Power_limit}
    P_{l}^{\text{limit}} = \frac{y}{x}.P_l^{\max} \\
\label{Eqn::Hardware_count_per_rack_group_xN_over_y} 
        N_{l, pod} = \lceil \frac{ (1+sm). N_{\text{row,pod}}. N_{\text{rack,row}}. P_{rack}^{\max} }{P_{l}^{\text{limit}}} \rceil  \qquad  ; l \in \text{ \{CDU, PDU\}} \\ 
        \label{Eqn::Total_hardware_count_xn_over_y}
        N_l = \lceil \frac{P_{DC}^{\max}}{P_{rack}^{\max} .N_{\text{rack,row}}. N_{\text{row,pod}} }  \rceil . N_{l, pod}
  \qquad  ; l \in \text{ \{CDU, PDU\}}  
    \end{gather}
\end{subequations}

For datacenter-level hardware -- including chillers, dry coolers, evaporative cooling towers, UPSs, MSBs, and backup generators --, we estimate the number of units required to support the IT load by dividing the datacenter's peak power demand (augmented by the safety margin) by the deliverable capacity of each unit, as shown in Equation~\ref{Eqn::Equipment_count_DC_level_IT_xN_over_y}. The same approach is applied to power delivery equipment serving the cooling system, in Equation~\ref{Eqn::Equipment_count_DC_level_Facility_xN_over_y}.

Finally, gray space utilization -- representing the area occupied by cooling and power infrastructure -- is calculated by summing the physical surface area ($A_l$) and the additional space required for maintenance access (modeled as a proportional factor $\lambda_{l}$) for each IT and facility hardware component.

\begin{subequations}
    \begin{gather}
\label{Eqn::Equipment_count_DC_level_IT_xN_over_y}
       N_{l,\text{IT}} = \lceil \frac{ (1+sm) . P_{DC}^{\max} }{P_l^{\text{limit}}} \rceil
      \qquad  ; l \in \text{ \{chiller, cooler, UPS, MSB, Gen\}} \\
\label{Eqn::Equipment_count_DC_level_Facility_xN_over_y}
        \resizebox{0.35\linewidth}{!}{$N_{l, \text{Facility}} = \lceil \frac{ (1+sm). \sum_k P_{k}^{\max} }{P_l^{\text{limit}}} \rceil $}
  \quad  ; \resizebox{0.45\linewidth}{!}{$l \in \text{ \{UPS, MSB, Gen\}} , k \in  \{CDU, chiller, cooler\} $} \\
  \label{Eqn::gray_space_}
  \text{Gray space}=  \sum_l (1+ \lambda_{l}) (N_{l,\text{IT}} +N_{l,\text{Facility}} ) . A_l    
    \end{gather}
\end{subequations}

\section{Case Studies}
\label{Sec:CaseStudies}
To illustrate the capabilities of DCGen, we exercise it to generate datacenter designs first defined by compute capability target and then based on a power target.  We then show how datacenter designs scale over time, using extrapolated technology capabilities to project to 2027 and 2029.

\subsection{Example of Single Datacenters based on xAI COLOSSUS}
\label{sec:SingleDCConfig}
In this case study, we examine how DCGen generates datacenter hardware configuration, employing the xAI Colossus system as a reference. 
We consider two cases: (1) the xAI Colossus datacenter which contains $1563 \times 100kW$ racks of GPU nodes and $203 \times 34kW$  racks of storage nodes , (2) a datacenter with twice the xAI Colossus compute capability (rack count). The study considers 42U racks.

Figure~\ref{fig:PowerDensityxAI}, Figure~\ref{fig:PowerCapacityxAI} and Figure~\ref{fig:FloorSpacexAI} show the power density, electrical power demand, and space utilization of both xAI colossus and the datacenter with twice its compute capability (racks count), respectively. 
The power density is not impacted by the datacenters scale. In fact, both cases are based on the same hardware configuration, hence the same amount of electrical power used by unit of space. 
Moreover, the hardware in the datacenter with twice xAI Colossus compute capability is scaled proportionally compared to the reference xAI Colossus datacenter. Therefore, the resulting electrical power demand and space utilization are scaled up proportionally, and double in comparison with the reference configuration. 

\begin{figure}[H]
  \centering
  \begin{subfigure}[b]{0.32\textwidth}
    \centering
     \includegraphics[width=\linewidth]{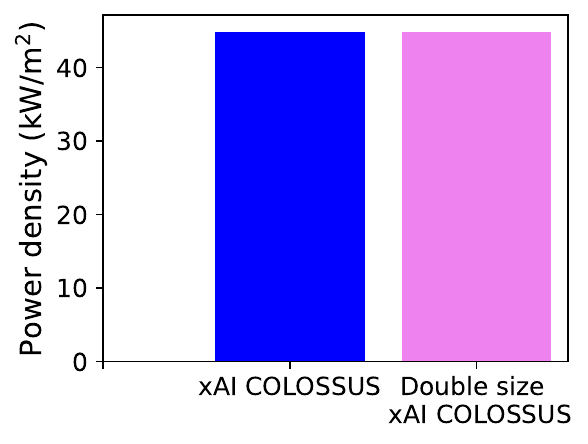}
    \caption{Power density \vspace{1em} }
    \label{fig:PowerDensityxAI}
  \end{subfigure}
  ~
  \begin{subfigure}[b]{0.32\textwidth}
    \centering
     \includegraphics[width=\linewidth]{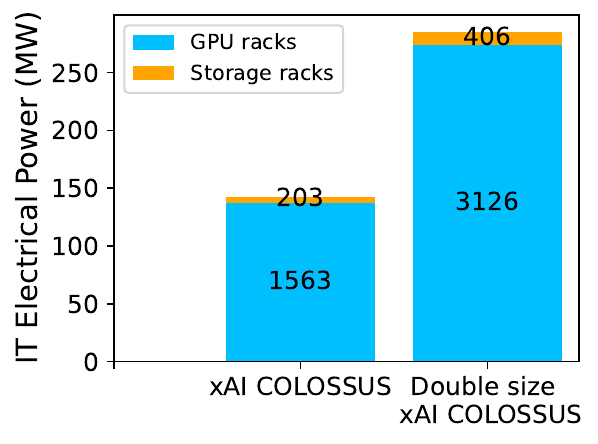}
    \caption{Electrical power. Rack count per node type shown in the bars}
    \label{fig:PowerCapacityxAI}
  \end{subfigure}
  ~
  \begin{subfigure}[b]{0.32\textwidth}
    \centering
     \includegraphics[width=\linewidth]{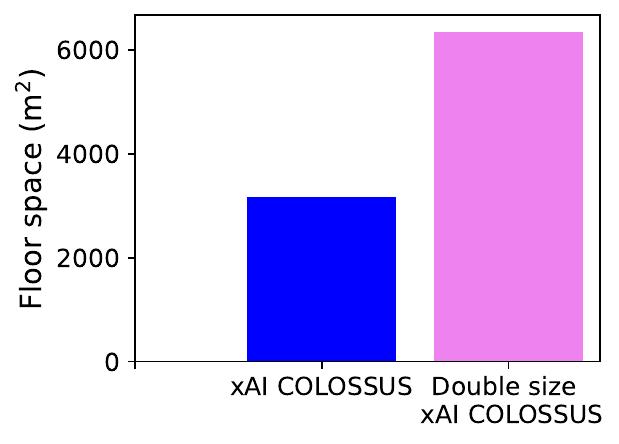}
    \caption{Space utilization \vspace{1.5em} }
    \label{fig:FloorSpacexAI}
  \end{subfigure}
  \caption{Single datacenter configurations. GPU racks and storage racks stand for racks containing GPU and storage nodes, respectively.}
  \label{fig:singleDataCenter}
\end{figure}


\subsection{ Datacenters defined by Compute Capability}
\label{Sec:DC-Example}
A fundamental consideration in datacenter hardware design is understanding the power used by one configuration or the other, within the same space. Hence, we analyze the power density and the electrical power of 10,000-rack datacenters, using the canonical models.

\begin{figure}[H]
  \centering
  \begin{subfigure}[b]{\linewidth}
    \centering
     \includegraphics[width=\linewidth]{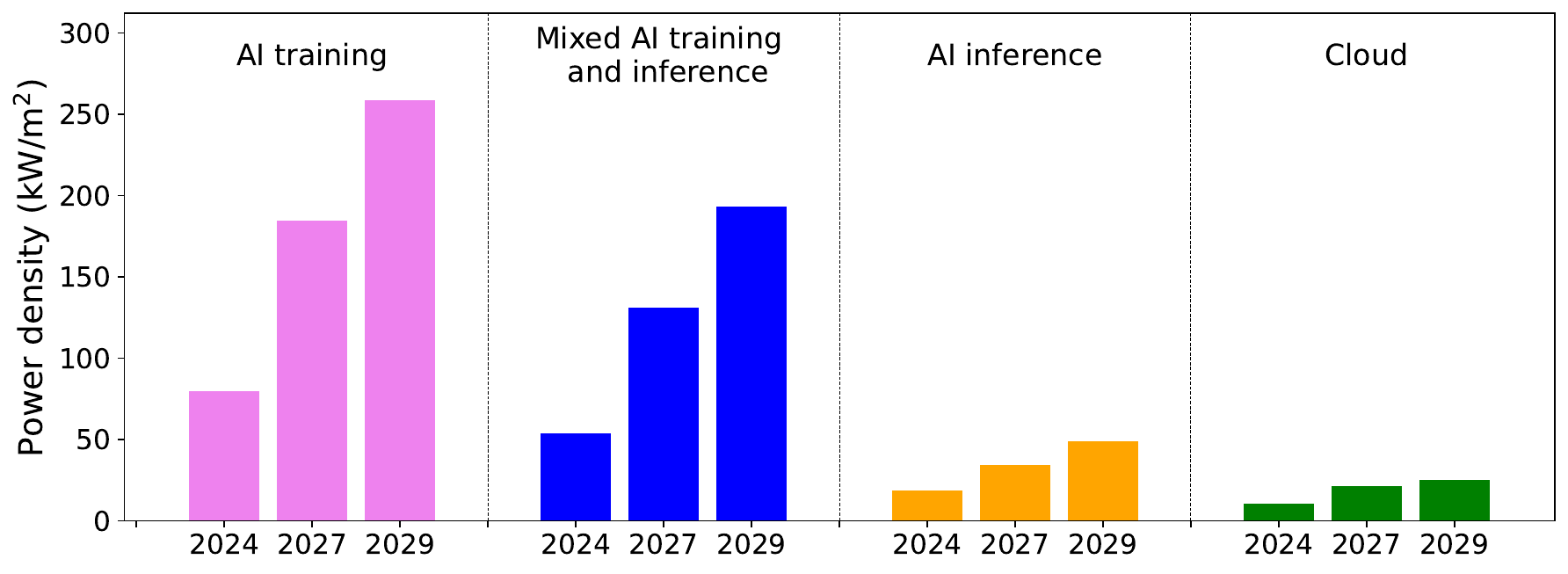}
    \caption{Power density}
    \label{fig:powerDensity1000racks}
  \end{subfigure}
  ~
  \begin{subfigure}[b]{\linewidth}
    \centering
     \includegraphics[width=\linewidth]{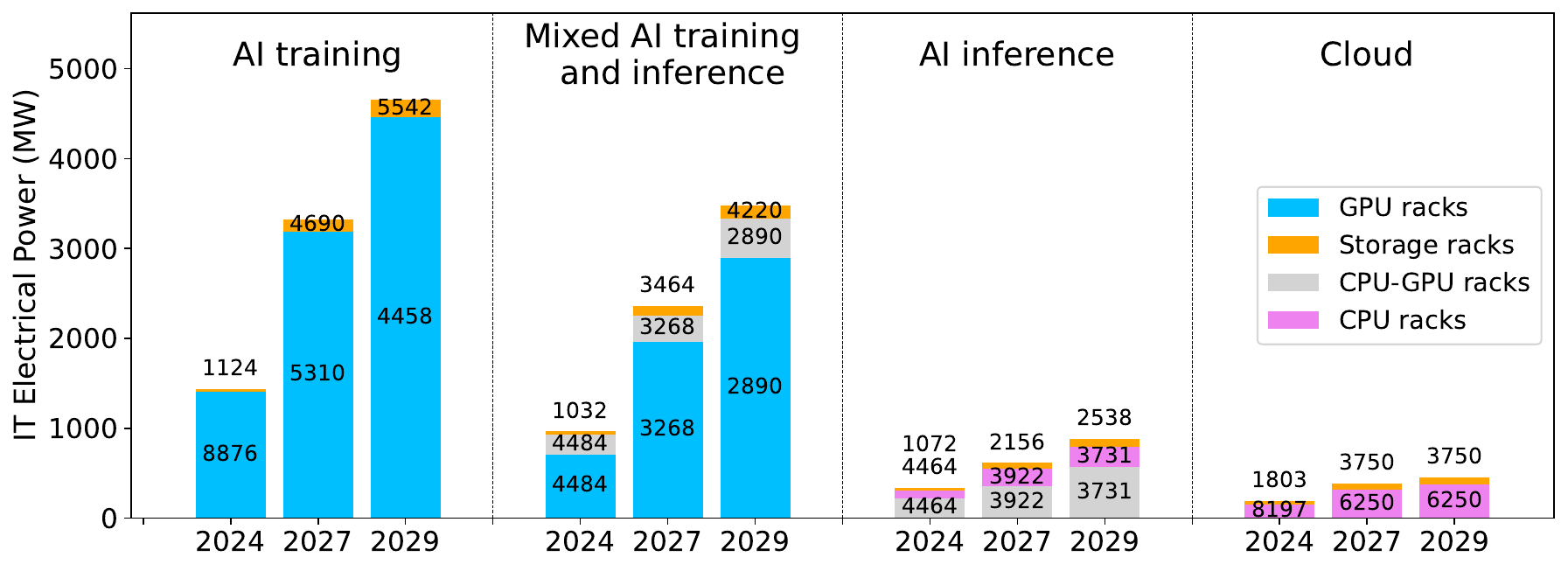}
    \caption{Electrical power. The number of racks are shown in the bars. }
    \label{fig:PowerCapacity1000racks}
  \end{subfigure}
  \caption{10,000-rack datacenters IT configurations.}
  \label{fig:1000rackDC}
\end{figure}

Figure~\ref{fig:powerDensity1000racks} and Figure~\ref{fig:PowerCapacity1000racks} show respectively the power density and the electrical power of four datacenter types (presented from left to right).  Today's  datacenters (2024) show highly variable power densities: 79.8 kW/m$^2$ for AI training, 53.7 kW/m$^2$ for mixed AI training and inference, 18.8 kW/m$^2$ for AI inference, and 10.4 kW/m$^2$ for conventional Cloud (Figure~\ref{fig:powerDensity1000racks}). This corresponds to total datacenter loads of 1.4GW, 963.3MW, 338.5MW, and 186.5MW respectively (Figure~\ref{fig:PowerCapacity1000racks}). In practical terms, AI training datacenters can pack 4.2$\times$ the power of AI inference and 7.7$\times$ that of Cloud datacenters into the same area.
Looking ahead, power density is expected to increase across all datacenter types. By 2027, AI training datacenters will be 2.3$\times$ denser, mixed AI training/inference 2.4$\times$, AI inference 1.8$\times$, and Cloud 2$\times$ denser than today. This growth will bring the total loads of these 10,000-rack scale datacenters to 3.3GW, 2.4GW, 613.1MW, and 381.5MW, respectively in 2027. Power density and power demand further grow in 2029. Power density is projected to 258.5kW/m$^2$ (3.2$\times$ the density of 2024) in AI training, 193.3kW/m$^2$ (3.6$\times$ 2024) in Mixed AI training and inference, 48.9kW/m$^2$ (2.6$\times$ 2024) in AI inference, and 25.2kW/m$^2$ (2.4$\times$ 2024) in Cloud datacenters.
This growth pushes the total loads of these 10,000-rack scale datacenters to 4.7GW, 3.5GW, 880MW, and 453.4MW.

\begin{figure}[H]
    \centering
   \begin{subfigure}[b]{\linewidth}
    \centering
      \includegraphics[width=\linewidth]{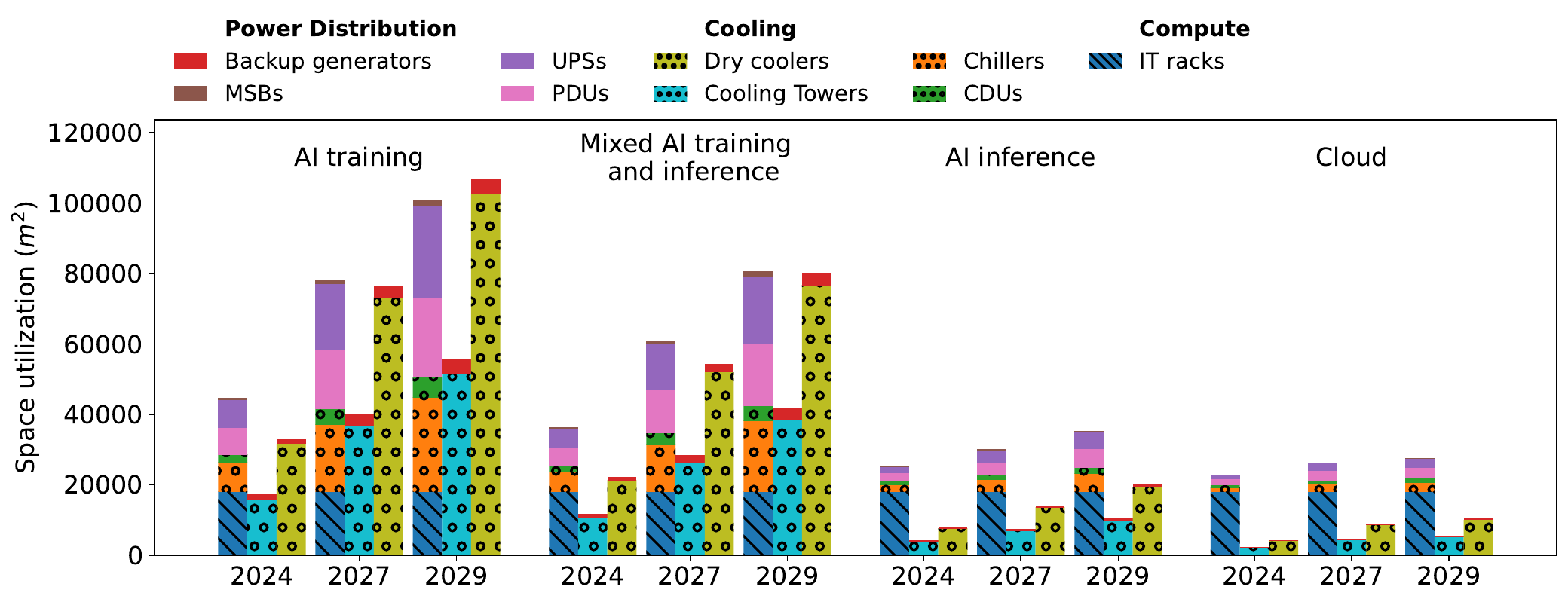}
    \caption{IT and Gray Space in Space-optimized design: indoor (first bars),  outdoor space in evaporative cooling (second bars), outdoor space in dry cooling (third bars).}
    \label{fig:Space1000racks}
  \end{subfigure}
   \begin{subfigure}[b]{\linewidth}
    \centering
      \includegraphics[width=\linewidth]{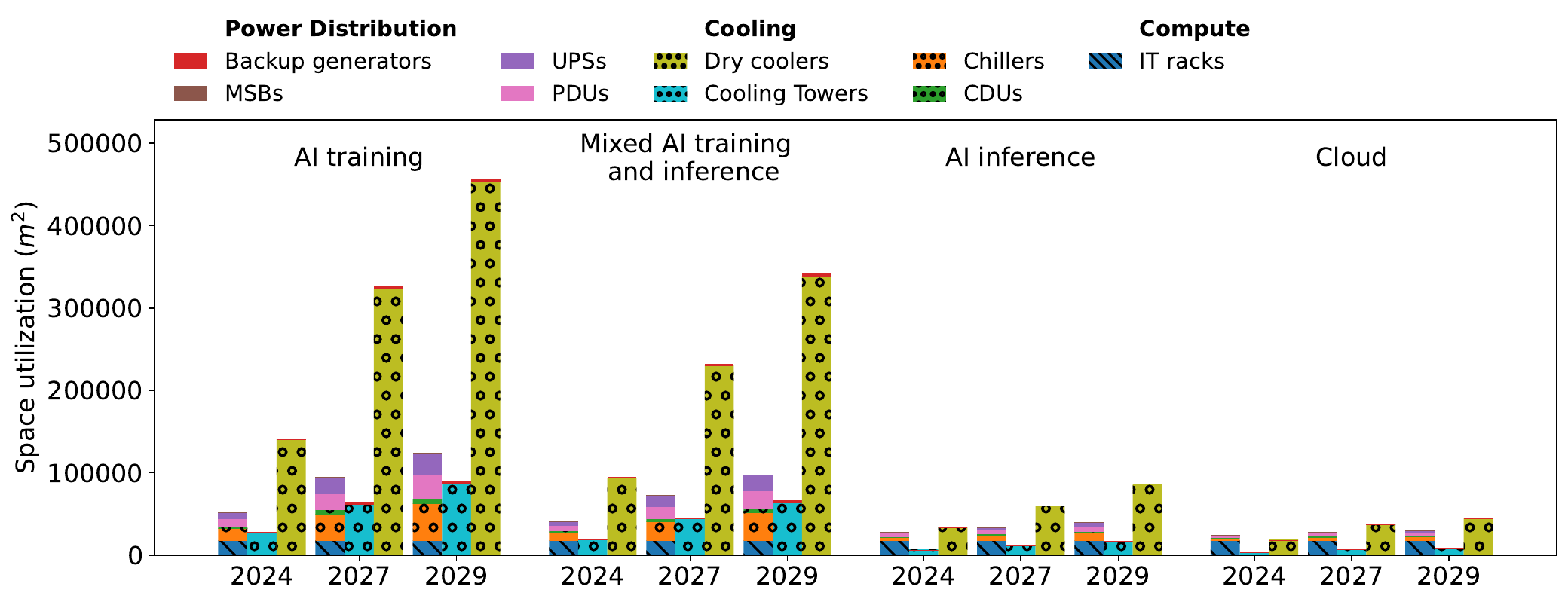}
    \caption{IT and Gray Space in Power-optimized design: indoor (first bars),  outdoor space in evaporative cooling (second bars), outdoor space in dry cooling (third bars).}
    \label{fig:Space1000racks_Power}
  \end{subfigure}
    \caption{DCGen 1.1 modeling of IT and Gray Space \added{for 10,000-rack datacenters.}} 
    \label{fig:Case-Study}
\end{figure}

Figure~\ref{fig:Space1000racks} shows space utilization for a space-optimized datacenter design, broken down by hardware emplacement: datacenter floor, outside with evaporative cooling, and outside with dry cooling. IT equipment consistently occupies 18,000m$^2$ across all datacenter types.
Indoor Gray space~\footnote{This estimated space reflects the minimum area needed for IT, cooling, and power infrastructure, not the total building footprint. For example, each building in Stargate's Abilene datacenter has a total footprint of 490,000~ft$^2$ (45,522~m$^2$)~\cite{crusoe_abilene_2025}, of which 35,000~ft$^2$ (3,252~m$^2$), or 7.1\%, is reported as white space across four data halls~\cite{dpr_crusoe_abilene}. The remaining 455,000~ft$^2$ (42,270~m$^2$) of indoor gray space is approximately 500\% larger than our space-optimized estimate for cooling and power infrastructure. Given the reported 100~MW IT capacity and 50,000 NVIDIA GB200 NVL72s per building, the implied rack deployment is substantially below the stated 2,000-rack capacity~\cite{dpr_crusoe_abilene}, suggesting significant unused white space. From this observation, we believe that indoor gray space is likewise also largely underutilized. A similar pattern appears at xAI's Colossus facility in Memphis, which was repurposed from a former Electrolux factory and expanded from 100,000 (2024) to 200,000 (2025) GPUs,~\cite{measuredai_xai_colossus} indicating that existing building size can provide substantial room for deployment and future expansion. We therefore encourage readers to distinguish between the minimal space requirements estimated by DCGen and the actual building footprints observed in industry practice.}  can be substantial, reaching 1.5$\times$ IT space in 2024 (AI training). In AI inference and Cloud datacenters, IT equipment dominates due to lower power density. By 2029, AI training  Gray space grows to 4.6$\times$ IT space in datacenter floor and 6.2$\times$ IT space outside (with dry coolers). Growth in Gray space has a smaller impact on AI inference and Cloud datacenters. 

Results for power-optimized design are shown in Figure~\ref{fig:Space1000racks_Power}. Gray space requirements becomes more substantial, especially outdoor. For instance, in 2024 datacenter configurations using dry coolers,  outdoor gray space represents  7.8$\times$ IT space for AI training, 5.3$\times$ IT space in mixed AI training and inference, 1.85$\times$ IT space in AI inference, and 1$\times$ IT space in Cloud datacenter. This space increases by 2.3$\times$ (2027) and  3.2$\times$ (2029) in AI training datacenter, 2.4$\times$ (2027) and  3.6$\times$ (2029) in Mixed AI training and inference, 1.8$\times$ (2027) and  2.6$\times$ (2029) in AI inference, and 2$\times$ (2027) and  2.4$\times$ (2029) in Cloud datacenter.

\subsection{Datacenter defined by Electrical Power}

Another datacenter design approach is to establish a target electrical power and then determine the number of hardware units and the space utilized for different use cases or workloads.
In this section, we compare the hardware setups of 1GW datacenters.

\begin{figure}[H]
  \centering
     \includegraphics[width=\linewidth]{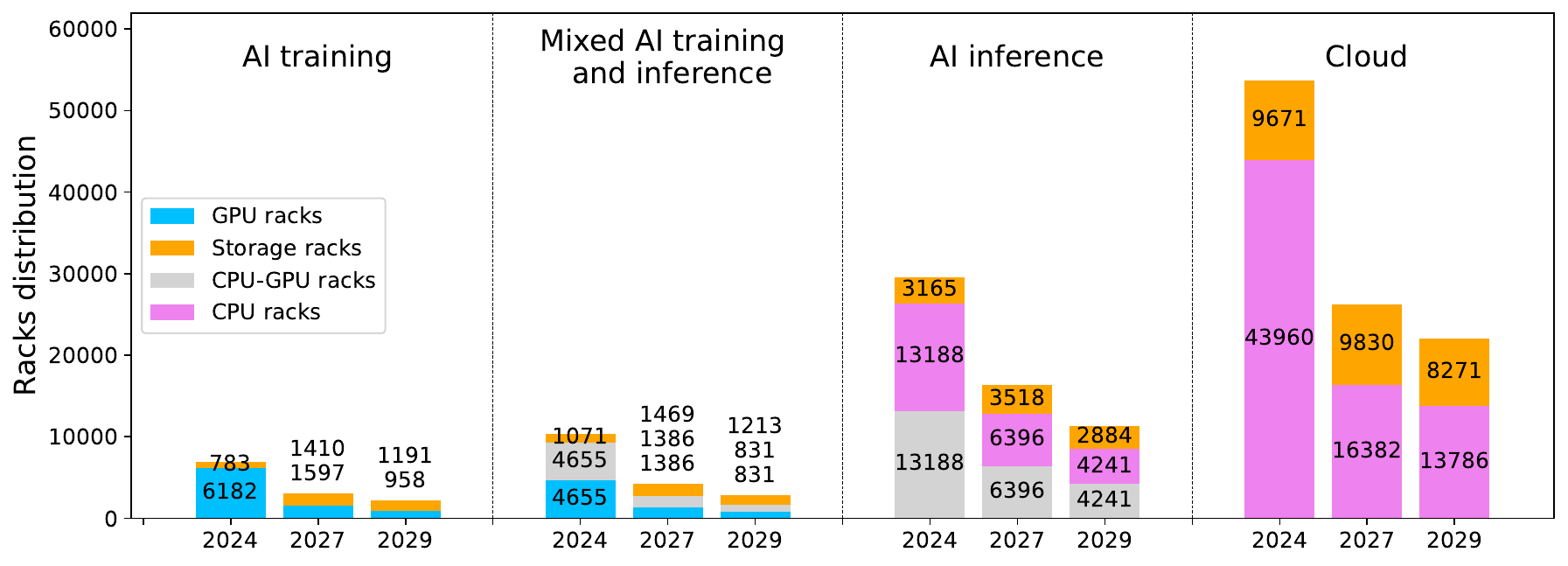}\\[-5pt]
    \caption{1GW datacenters IT configuration.}
    \label{fig:RackDisto1GW}
  \end{figure}

Figure~\ref{fig:RackDisto1GW} and~Figure\ref{fig:1GWDC} present respectively the IT racks allocation per node types and space utilization in the four datacenters. 
Today's 1GW-scale AI training datacenter is made of 6,965 (42U) racks. Other datacenter types require significantly more racks: 1.5$\times$ for mixed AI training/inference, 4.2$\times$ for AI inference, and 7.7$\times$ for Cloud (Figure~\ref{fig:RackDisto1GW}). 
Future datacenters will need fewer compute racks to deliver the same power load (Figure~\ref{fig:RackDisto1GW} and Figure~\ref{fig:SPACE1GW}). We project rack counts to reduce by 56.8\% in AI training, 59.1\% in mixed AI training and inference, 44.7\% in AI inference, and 51.1\% in Cloud datacenters, with this downward trend continuing in 2029. Gray space (see Figure~\ref{fig:SPACE1GW}) remains nearly unchanged, as most of the cooling and power distribution equipment are located at the datacenter-level.

\begin{figure}[H]
  \centering
  \begin{subfigure}{\linewidth}
    \centering
      \includegraphics[width=\linewidth]{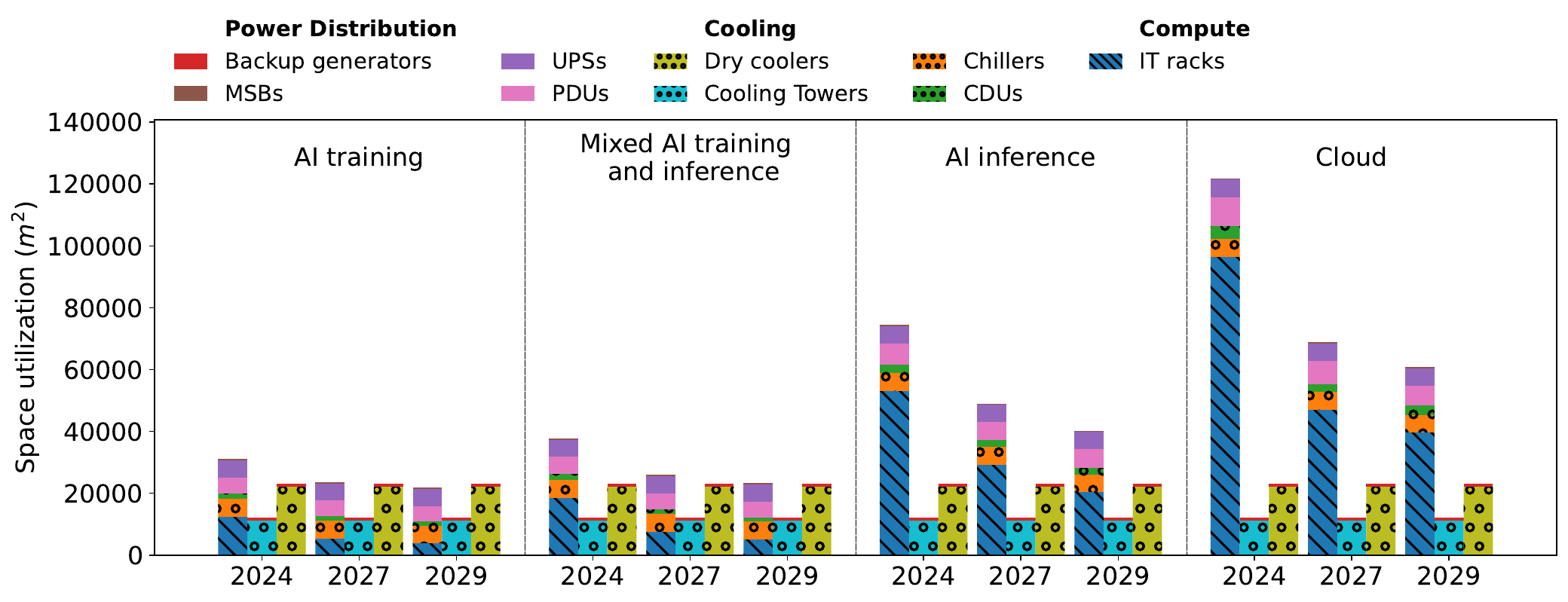}
    \caption{IT and Gray Space in Space-optimized design: indoor (first bars),  outdoor space in evaporative cooling (second bars), outdoor space in dry cooling (third bars).}
    \label{fig:SPACE1GW}
  \end{subfigure}
  
  \begin{subfigure}{\linewidth}
    \centering
      \includegraphics[width=\linewidth]{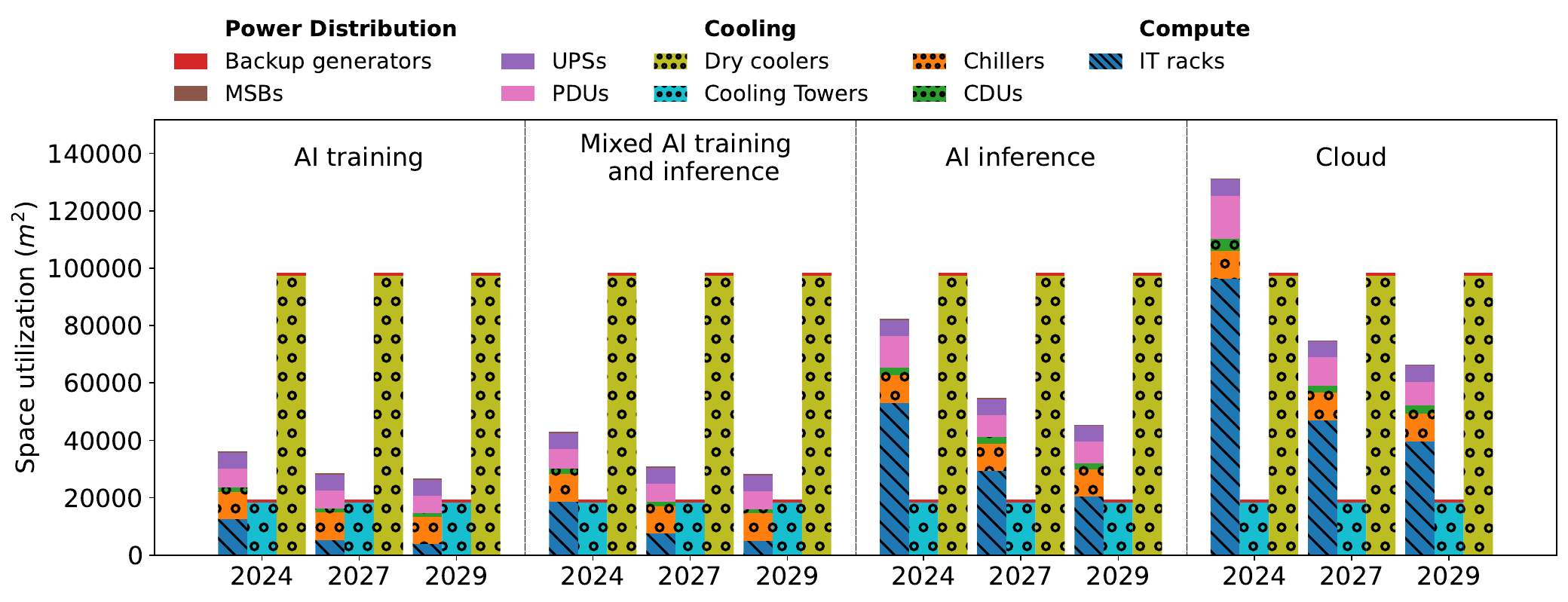}
    \caption{IT and Gray Space in Power-optimized design: indoor (first bars), outdoor space in evaporative cooling (second
bars), outdoor space in dry cooling (third bars).}
    \label{fig:Space1GW_Power}
  \end{subfigure}
  \\[-10pt]
  \caption{\deleted{1GW datacenters space utilization.} \added{DCGen 1.1 modeling of IT and Gray Space for 1GW datacenters.}} 
  \label{fig:1GWDC}
\end{figure}

\section{Conclusion}
DCGen is a model-driven tool for generating realistic datacenter configurations, including IT, cooling, and power systems. Its current \deleted{1.0} \added{1.1} version generates designs (2024, 2027, 2029) for four types of datacenters \added{(AI Training, Mixed AI Training and Inference, AI Inference, Cloud)}, with compute or power targets.  DCGen also provides key metrics like power density and space utilization for in-depth analysis, e.g., future-proofing designs, reliability and redundancy studies, and sustainability efforts. \added{The tool is available as open source at this link\textsuperscript{\ref{DCGen-git-link}}}.

\section*{Acknowledgements}

Funding for this work was provided by the U.S. Department of Energy (DOE), Office of Energy Efficiency and Renewable Energy Geothermal Technologies Office, and thru 
the
\deleted{National Renewable Energy Laboratory (NREL)} \added{National Laboratory of the Rockies (NLR)} under Contract No. DE-AC36-08GO28308.
The views expressed in the article do not necessarily represent the views of the DOE or the U.S. Government. 

\printbibliography
\end{document}